\documentclass[12pt]{article}
\usepackage{latexsym}
\usepackage{epsfig}
\usepackage{amssymb}
\usepackage{amsthm}
\usepackage{amsmath}
\usepackage{amssymb,amsfonts}
\usepackage{graphicx}
\usepackage{verbatim}

\newcommand{\be}{\begin{equation}}
\newcommand{\ee}{\end{equation}}
\newcommand{\bea}{\begin{eqnarray}}
\newcommand{\eea}{\end{eqnarray}}
\newcommand{\nn}{\nonumber}
\newcommand{\C}{\mathcal{C}}

\numberwithin{equation}{section}

\DeclareMathOperator{\tr}{\textrm{tr}}
\DeclareMathOperator{\Tr}{\textrm{Tr}}

\DeclareMathOperator{\diag}{\textrm{diag}}

\begin{document}
\pagestyle{empty}
\centerline{ \hfill hep-th/0407221 }
\vspace{3.5cm}

\begin{center}
{\Large \bf Gauged $\rm \bf D=7$ Supergravity on the $\rm \bf
S^1/\mathbb{Z}_2$ Orbifold}
\end{center}
\vspace{1.cm}

\centerline{\bf Spyros D. Avramis and Alex Kehagias }

\vspace{.3cm}

\centerline{\rm \small Department of Physics, Ethniko Metsovio
Polytechnio, GR-15773, Zografou, Athens, Greece}

\vspace{.5cm}

\centerline{\em e-mail : avramis@cern.ch,~kehagias@cern.ch}
\vspace{2.cm}

\centerline{ \bf Abstract}

\begin{quote}

\indent

We construct the most general couplings of a bulk
seven-dimensional Yang-Mills-Einstein ${\cal{N}}=2$ supergravity
with a boundary six-dimensional chiral ${\cal{N}}=(0,1)$ theory of
vectors and charged hypermultiplets. The boundary consists of two
brane worlds sitting at the fixed points of an ${\bf
S}^1/\mathbb{Z}_2$ compactification of the seven-dimensional bulk
supergravity. The resulting 6D massless spectrum surviving the
orbifold projection is anomalous. By introducing boundary fields
at the orbifold fixed points, we show that all anomalies are
cancelled by a Green-Schwarz mechanism. In addition, all couplings
of the boundary fields to the bulk are completely specified by
supersymmetry. We emphasize that there is no bulk Chern-Simons
term to cancel the anomalies. The latter is traded for a
Green-Schwarz term which emerges in the boundary theory after a
duality transformation implemented to construct the bulk
supergravity.

\end{quote}
\vspace{1cm}

\newpage
\tableofcontents

\newpage

\newpage

\eject \pagestyle{empty} \setcounter{page}{1}
\setcounter{footnote}{0}
\pagestyle{plain}
\pagenumbering{arabic}

\section{Introduction}
\label{s-1}

The idea that we live on a brane in a higher-dimensional spacetime
has led recently to new possibilities for physics beyond the
standard model. In the brane world, where the geometry of extra
dimensions can naturally account for
hierarchies~\cite{Arkani-Hamed:2001is,Arkani-Hamed:1998rs,
Randall:1999ee,Randall:1999vf}, one  hopes to find constraints
which will specify the matter content on the brane. For example,
we know that the cancellation of anomalies is a crucial guiding
principle for the construction of consistent theories. We are
familiar with the  cancellation of gauge anomalies in four
dimensions, whereas anomaly cancellation in higher
dimensions~\cite{Scrucca:2004jn} may also lead to powerful
constraints, especially in those circumstances where gravitational
and mixed anomalies on top of gauge ones are present as in the
ten-dimensional heterotic/Type I and Type IIB string theories. In
particular, anomaly cancellation has turned out to be of great
significance in the context of effective field theories arising as
low-energy limits of fundamental theories whose detailed
high-energy structure is not known. In such a case, anomaly
cancellation conditions enable us to infer information about the
high-energy theory by studying low-energy physics.

In the brane world context, the prototype model combining the
above ideas has been constructed by Ho\v rava and Witten
\cite{Horava:1995qa,Horava:1996ma}. The goal was to provide an
11-dimensional interpretation of the $E_8 \times E_8$ heterotic
string. To obtain a chiral $D=10$ spectrum from M-theory,
compactification on the ${\bf S}^1/\mathbb{Z}_2$ orbifold had been
considered. The fixed points of the $\mathbb{Z}_2$ action model
two brane worlds of zero tension. The orbifold projection
eliminates certain bosonic fields and imposes chirality conditions
on the spinors, thereby resulting in a $D=10$ chiral spectrum.
However, the price to pay is the appearance of anomalies, whose
cancellation requires the presence of extra boundary fields, i.e.
vector multiplets, which are the only ones available in ten
dimensions. The anomaly cancellation conditions uniquely determine
the gauge group on each fixed plane to be $E_8$, and anomalies are
cancelled by inflow from Chern-Simons and Green-Schwarz terms.
Moreover, the gauge coupling constant which seems to enter as a
new free parameter in the theory turns out to be related to the
gravitational coupling. Finally, the dynamics of the resulting
theory in the limit of small compactification radius are believed
to correspond to a strongly coupled version of the $E_8 \times
E_8$ heterotic string. The work of Ho\v rava and Witten provided a
missing link in the impressive chain of string dualities and
triggered much recent interest in brane-world scenarios.

Anomaly cancellation then on brane worlds in eleven dimensions
severely constrains, and in fact, completely specifies the gauge
structure.  This should not be anticipated for the brane worlds
constructed in five
dimensions~\cite{Kehagias:1999ju},\cite{Kehagias:2000au}, where
the boundary gauge group is not restricted by any local anomaly,
although global anomalies may impose some
constraints~\cite{GrootNibbelink:2002qp,Gmeiner:2002ab,Gmeiner:2002es}.
However, gravitational anomalies also exist in six
dimensions~\cite{Alvarez-Gaume:1983ig}, and this places nontrivial
constraints on six-dimensional (6D)
theories~\cite{Salam:1985mi}-\cite{Nibbelink:2003rc}. Therefore,
it is expected that the matter on six-dimensional brane worlds in
seven dimensions will also be restricted, and may even be uniquely
specified,  by anomaly cancellation. This expectation has been
studied in~\cite{Gherghetta:2002xf,Gherghetta:2002nq} where it was
shown that in seven-dimensional (7D) brane worlds the gauge group
structure and matter content on the boundaries will be similarly
restricted. Of course, the dimensional reduction of the Ho\v
rava-Witten (HW) model automatically gives rise to brane worlds
that are
anomaly-free~\cite{Gorbatov:2001pw,Kaplunovsky:1999ia,Faux:2000sp}.
However, this is not the only possibility. The starting point of
the construction of~\cite{Gherghetta:2002xf,Gherghetta:2002nq},
which we summarize, is ${\cal N}=2$ 7D gauged supergravity with a
3-form potential. The ungauged theory is obtained from the
compactification of M-theory on K3 or, equivalently, from the
compactification of strongly coupled heterotic theory on
$T^3$~\cite{Witten:1995ex}. The compactification produces twenty
two vectors resulting from expanding the eleven-dimensional
3--form on the $b_2=22$ 2--cycles of the K3. Three of these
vectors are members of the gravity multiplet, whereas the
remaining nineteen fill vector multiplets of the ${\cal N}=2$ 7D
theory. Each vector multiplet also contains three scalars, and the
57 total scalars parameterize the coset space
$SO(19,3)/SO(19)\times SO(3)$, for which an $SO(3)\times H$ or
$SO(3,1)\times H$ subgroup of $SO(19,3)$ can be gauged. A
supersymmetric gauged theory can be obtained after introducing an
appropriate potential for the scalar field (corresponding to the
K3 volume). The scalar potential has two extrema, leading to
either a supersymmetric or non-supersymmetric
vacuum~\cite{Mezincescu:ta}. The supersymmetric vacuum has a
negative cosmological constant implying that the vacuum in the
gauged theory is not Minkowski spacetime but rather anti-de
Sitter, $AdS_7$. The $AdS_7$ vacuum with ${\cal N}=2$
supersymmetry has been considered in the context of the AdS/CFT
correspondence \cite{Maldacena:1997re}, and was shown to be the
supergravity dual of the 6D ${\cal N}=(1,0)$
SCFT~\cite{Ferrara:1998vf},\cite{Kehagias:1998gn}.

The minimal ${\cal N}=2$ 7D gauged supergravity may be
compactified down to six dimensions on ${\bf S}^1$, even in the
presence of a cosmological constant as was shown
in~\cite{Giani:dw}, resulting in a non-chiral ${\cal N}=(1,1)$ 6D
theory. However, we are interested in the chiral ${\cal N}=(0,1)$
6D theory because in this case vector, tensor and hypermultiplets
can couple to gravity in a way that is restricted by anomalies. In
particular, the possibility of vector multiplets on the 6D
boundaries allows one to construct theories which contain the
standard model gauge group. Thus, we need to find a way to obtain
the chiral 6D theory from the 7D gauged supergravity.

An immediate way to obtain the 6D chiral theory is to compactify
on an orbifold ${\bf S}^1/\mathbb{Z}_2$. Besides the localized
gravity multiplet there will also be a localized tensor and
hypermultiplet in the resulting 6D ${\cal N}=(0,1)$ chiral theory.
However,  the spectrum found  was anomalous because in six
dimensions there are gravitational anomalies. These anomalies were
then cancelled by appropriate boundary fields such as vector,
tensor and hypermultiplets. The locally supersymmetric
bulk-boundary couplings were derived for the case of boundary
vector and {\it neutral} hypermultiplets. In the HW model, the
Bianchi identity for the four-form field strength had to be
modified in order to obtain a consistent coupling between the
boundary gauge couplings and the bulk. In the scenario
of~\cite{Gherghetta:2002xf,Gherghetta:2002nq} described above,
where only the gravity multiplet was considered in the bulk and
the boundary fields were restricted to  {\it neutral}
hypermultiplets, a similar modification for the Bianchi identity
occurs as well.

In this paper, we construct the gauged $D=7$ minimal
$\mathcal{N}=2$ supergravity on a manifold with boundaries. The
boundaries are the fixed points of the  ${\bf S}^1/\mathbb{Z}_2$
orbifold compactification of the 7D supergravity. The starting
point is the 7D minimal $\mathcal{N}=2$ supergravity theory
containing the gravity multiplet and $N$ vector multiplets, gauged
with respect to a subgroup of the holonomy group $SO(N) \times
SO(3)$ of the scalar manifold. There are two versions of this
theory. The first one is based on the gravity multiplet with the
3-form potential and has been constructed in
\cite{Mezincescu:ta,Townsend:1983kk,Giani:dw}. This is also the
case consider in~\cite{Gherghetta:2002xf,Gherghetta:2002nq} with
no vectors however. There is also the possibility for the 3-form
to be traded for a 2-form (since 2-and 3-forms have field
strengths which are Poincar\'e duals in 7D) and the 2-form version
of the minimal 7D supergravity has been constructed in
\cite{Salam:1983fa,Han:1985ku}. We present the version of the
theory with a 3--form potential $A_3$ which can be dualized (in
the absence of a topological mass term) to an equivalent theory
phrased in terms of a 2--form potential $B_2$. The dualization can
be performed according to familiar methods but in our case, care
should be given to boundary terms occurring during the dualization
procedure. We need to retain such boundary terms, which are
usually ignored, as in our case we have boundaries. These boundary
contributions give rise to a Green-Schwarz term, necessary for the
cancellation of anomalies.

The dimensional reduction of the resulting  2--form version of the
theory on the ${\bf S}^1/\mathbb{Z}_2$ orbifold entails the
assignment of a $\mathbb{Z}_2$ parity on the various fields as
dictated by invariance of the action and supersymmetry. After
modding out $\mathbb{Z}_2$, the odd fields are projected out and
the surviving fields fit into multiplets of chiral $D=6$,
$\mathcal{N}=(0,1)$ supergravity. The boundary theory arising from
dimensional reduction can be constructed in a straightforward way
and it is what one obtains by truncating $D=6$, $\mathcal{N}=4$
supergravity according to the $\mathbb{Z}_2$ projection.

On the other hand, the resulting theory is not self-consistent at
this stage. First, the supersymmetry transformation laws contain a
certain term that breaks local supersymmetry on the fixed planes.
The resulting variations can be cancelled by adding certain
boundary potential terms for the scalars, which are interpreted as
brane tensions. These terms enter with opposite signs for the two
branes and therefore, we have a positive- and a negative-tension
brane. Second, since the $D=6$ theories on the fixed planes have a
chiral spectrum, they also have gravitational anomalies. To cancel
these anomalies, one must follow the Ho\v rava-Witten recipe and
introduce additional vector multiplets, tensor multiplets and
hypermultiplets living on the fixed planes. As illustrated
in~\cite{Gherghetta:2002xf,Gherghetta:2002nq} for the
$Sp(1)$--gauged theory, there are many possibilities regarding the
choice of the boundary gauge groups. The anomaly cancellation
conditions are similar to those presented there and can be solved
for many choices of the boundary gauge group.

Our final task is to construct the action of the boundary theory.
We will consider the case where (i) no additional boundary tensor
multiplets are introduced, (ii) the gauge group contains the
$Sp(n_H)\times Sp(1)$ holonomy group of the hyperscalar manifold
and (iii) the boundary hypermultiplets are \emph{charged} under
the gauge group. Starting from the vector multiplets, the
appropriate action can be constructed along the lines
of~\cite{Horava:1995qa,Horava:1996ma} starting from the globally
supersymmetric theory and introducing additional couplings with
the bulk fields. In particular, as in the HW case, local
supersymmetry dictates that the 7-component of the 3--form field
strength tensor $G_3$ should acquire a boundary value proportional
to $*\tr F^2$, where $F$ is the Yang-Mills field strength. This
introduces a $B_2 \wedge \tr F^2$ interaction, which is well-known
to exist in $D=6$ supergravity theories coupled to tensors and
plays the role of the Green-Schwarz term.

Passing on to the hypermultiplets, we can initially construct the
locally supersymmetric action for the case where they are inert
under the gauge group and then gauge the multiplets by defining
appropriate covariant derivatives. Consistency of the
supersymmetry transformations then demands that the covariant
derivative acting on the spinor $\epsilon$ parameterizing the SUSY
transformations should also involve the boundary gauge fields and
this in tern implies that the variation of the gravitino kinetic
term acquires an extra term. To cancel this term, we follow the
Noether method and we introduce additional couplings.

The outline of this paper is as follows. In Section 2 we present
the general formalism for the consistent coupling of 7D vector
multiplets to the graviton multiplet. The 3-form version of the 7D
minimal ${\cal{N}}=2$ supergravity  is presented and the 2-form
version is obtained by duality transformation keeping boundary
terms which will be used later. In section 3 we discuss the ${\bf
S}^1/\mathbb{Z}_2$ orbifold compactification of the 2-form version
and the resulting effective 6D theory. In section 4 we present
couple the bulk supergravity to the 6D boundary theory consisting
of vectors and charged hypermultiplets and no tensor multiplets.
Without loss of generality and for notational simplicity, the
$4n_H$ scalars of the hypermultiplet are taken to parameterize the
coset $Sp(n_H,1)/ Sp(n_H)\times Sp(1)$, while the gauge group
contains the $Sp(n_H)\times Sp(1)$ holonomy group of the
hyperscalar manifold. In Section 5 we consider the total anomaly
of the theory which is localized at the boundaries and explain the
anomaly cancellation through the Green-Schwarz mechanism. In
Section 6 we present the complete total bulk and boundary
Lagrangian invariant under supersymmetry. It is the most general
Lagrangian with no tensor multiplets on the boundary up to
four-fermi terms. Finally, in Appendix A, B we summarize our
notation and conventions.

\section{7-dimensional $\mathcal{N}=2$ supergravity}
\label{s-2}

Our starting point is the gauged $D=7$, $\mathcal{N}=2$ (minimal)
supergravity theory. The supergravity multiplet can be described
either in terms of a 3--form potential or in terms of its dual
2--form potential and has a $Sp(1)$ rigid symmetry that can be
gauged. The resulting theories were first constructed in
\cite{Salam:1983fa,Han:1985ku} (2--form version) and
\cite{Mezincescu:ta,Townsend:1983kk,Giani:dw} (3--form version).
The theory can also be coupled to $N$ vector multiplets, in which
case the scalars parameterize a coset space and a subgroup of its
$SO(N) \times SO(3)$ holonomy group can also be gauged. These
theories were constructed in \cite{Bergshoeff:1985mr,Park:id}.

In this section, starting with the 3--form version of the gauged
$D=7$, $\mathcal{N}=2$ supergravity, we construct the 2--form
version via a duality transformation of the standard type
\cite{Cremmer:1979up,Nicolai:1980td}. The difference in our case
is that, we allow the 7D manifold on which we perform the duality
transformation to have boundaries.  Then, in addition to standard
practice, we need also retain a certain boundary term that emerges
during the duality transformation. This term will come into play
when we will discuss anomaly cancellation in the 6D supergravity
living on the boundary of the 7D space.

\subsection{General formalism}

The field content of the massless representations of the minimal
$D=7$, $\mathcal{N}=2$ supersymmetry algebra consists of the
following multiplets
\begin{eqnarray}
\label{e-2-1}
\text{Gravity multiplet} \quad&:&\quad ( g_{MN}, A_{MNP},
A^{\phantom{M} i}_{M \phantom{i} j},
 \phi, \psi^i_M, \chi^i ), \nn\\
\text{Vector multiplet} \quad&:&\quad ( A_M, \phi^{i}_{\phantom{i}
j}, \theta^{i} ), 
\end{eqnarray}
where all spinors are symplectic Majorana and the index $i=1,2$
takes its values from the $Sp(1)$ R-symmetry group of the algebra.
Thus, the gravity multiplet contains the graviton $g_{MN}$, an
antisymmetric 3--form $A_{MNP}$, an $Sp(1)$ triplet of vectors
$A_{M \phantom{i} j}^{\phantom{M} i}$, a scalar $\phi$, the
gravitinos $\psi_M^i$ and spinors $\chi^i$, whereas the vector
multiplet contains a vector $A_M$, an $Sp(1)$ triplet of scalars
$\phi^{i}_{\phantom{i} j}$ and an $Sp(1)$ symplectic Majorana
spinor $\theta^i$. The $Sp(1)$ R-symmetry can be gauged and the
resulting ${\cal{N}}=2$ 7D gauged supergravity has been
constructed in \cite{Mezincescu:ta,Townsend:1983kk,Giani:dw}.

The coupling of $N$ vector multiplets to the 7D $\mathcal{N}=2$
supergravity leads to the reducible multiplet
\begin{equation}
{\cal{V}}_3=(g_{MN}, A_{MNP}, A_M^I, \phi^\alpha, \phi, \psi^i_M,
\chi^i, \theta^{ai} ), \label{e-2-2}
\end{equation}
where $a=1,\ldots,N$ labels the individual vector multiplets,
$I=1,\ldots,N+3$ labels the vector fields resulting from the
combination of $A^{\phantom{M} i}_{M \phantom{i} j}$ and $A^a_M$
and $\alpha=1,\ldots,3N$ labels the scalars which parameterize the
coset space $SO(N,3)/SO(N) \times SO(3)$. Trading the 3--form
$A_{MNP}$ for a dual 2-form $B_{MN}$, we obtain the reducible
multiplet
\begin{equation}
{\cal{V}}_2=(g_{MN}, B_{MN}, A_M^I, \phi^\alpha, \phi, \psi^i_M,
\chi^i, \theta^{ai} ). \label{e-2-3}
\end{equation}
The action of the theory was first found in
\cite{Bergshoeff:1985mr} and \cite{Park:id}, in the $2$--form and
$3$--form version respectively. An $(N+3)$--parameter subgroup of
the isometry group $SO(N,3)$ of the scalar manifold can be gauged.
An important consequence of this gauging is the emergence of a
scalar potential, which is however indefinite.

It is useful to assemble the $3N$ scalars into a $N \times 3$
matrix $\boldsymbol{\Phi}$ and define the $(N+3) \times (N+3)$
matrix $\mathbf{L}$ with components
\begin{equation}
L_I^{\phantom{I} A} = \left[ \exp \left( \begin{array}{cc} 0 & \boldsymbol{\Phi} \\
\boldsymbol{\Phi}^T & 0 \end{array} \right) \right]_I^{\phantom{I}
A},~~~~A,I=1,\cdots,N+3. \label{e-2-4}
\end{equation}
which satisfies the $SO(N,3)$ orthogonality condition
\begin{equation}
\eta_{AB} L_I^{\phantom{I} A} L_J^{\phantom{J} B} = \eta_{IJ}
\qquad;\qquad \eta_{AB} = \diag(-,-,-,+,\ldots,+). \label{e-2-5}
\end{equation}
One needs also consider the inverse matrix $\mathbf{L}^{-1}$ with
components given by
\begin{equation}
L^I_{\phantom{I} A} = \eta^{IJ} \eta_{AB} L_J^{\phantom{J} B},
\label{e-2-6}
\end{equation}
which satisfies
\begin{equation}
L_I^{\phantom{I} A} L^I_{\phantom{I} B} = \delta^A_{\phantom{A}
B}. \label{e-2-7}
\end{equation}
We may gauge now a subgroup of the isometry group $SO(N,3)$ of the
scalar manifold $SO(N,3)/SO(N) \times SO(3)$. For this, a subgroup
$G \subset SO(N,3)$ is needed, whose dimension should equal the
number of vectors of the theory, i.e. $\dim G = N+3$. Let
$f_{IJ}^{\phantom{IJ} K}$ be the structure constants of $G$ and
let $D = d + i A$ be the gauge-covariant derivative (here, we have
absorbed the gauge coupling(s) into the structure constants).
Then, from $\mathbf{L}$, we can construct the Maurer-Cartan form
$\mathbf{L}^{-1} D \mathbf{L}$ with components
\begin{equation}
L^I_{\phantom{I} A} D_M L_I^{\phantom{I} B} = L^I_{\phantom{I} A}
\left(  \partial_M \delta_I^{\phantom{I} K} + f_{IJ}^{\phantom{IJ}
K} A^J_M \right) L_K^{\phantom{I} B}. \label{e-2-8}
\end{equation}
For the construction of the action, it is convenient to decompose
the index $A$ in $L_I^{\phantom{I} A}$ in terms of $SO(N)$ and $SO(3) \cong SU(2)$ indices according to
\begin{equation}
L_I^{\phantom{I} A} = \left( L_{I}^{\phantom{I} n}, L_I^{\phantom{I} a} \right) = \left( L_{I \phantom{i} j}^{\phantom{I} i}, L_I^{\phantom{I} a} \right). \label{e-2-9}
\end{equation}
where $n$ and $i$ are triplet and doublet indices for $SU(2)$ respectively\footnote{Passing from $SU(2)$ triplet indices to doublet indices is accomplished by $L_{I \phantom{i} j}^{\phantom{I} i} = \frac{1}{\sqrt 2} L_I^{\phantom{I} m} (\sigma_m)^i_{\phantom{i} j}$ and $Q_{M \phantom{i} j}^{\phantom{M} i} = \frac{i}{2} \epsilon_{mnp} (\sigma^m)^i_{\phantom{i} j} Q_M^{\phantom{M} np}$ etc.}. The constraint (\ref{e-2-5}) then is written as
\begin{equation}
L_I^{\phantom{I} a} L_{Ja} - L_{I \phantom{i} j}^{\phantom{I} i}
L_{J \phantom{j} i}^{\phantom{J} j} = \eta_{IJ}, \label{ij}
\end{equation}
while the Maurer-Cartan form (\ref{e-2-8}) decomposes into components as 
\begin{equation}
\label{e-2-10x}
P_{Ma}^{\phantom{MA} n} = L^I_{\phantom{I} a} D_M L_{I}^{\phantom{I} n}, \qquad
Q_{Ma}^{\phantom{Ma} b} = L^I_{\phantom{I} a} D_M L_I^{\phantom{I} b}, \qquad
Q_{Mm}^{\phantom{Mm} n} = L^{I}_{\phantom{I} m} D_M L_{I \phantom{n}}^{\phantom{I} n}.
\end{equation}
or, employing the $SU(2)$ doublet notation as in \cite{Bergshoeff:1985mr},
\begin{equation}
\label{e-2-10}
P_{Maj}^{\phantom{MAj} i} = L^I_{\phantom{I} a} D_M L_{Ij}^{\phantom{Ij} i}, \qquad
Q_{Ma}^{\phantom{Ma} b} = L^I_{\phantom{I} a} D_M L_I^{\phantom{I} b}, \qquad
Q_{M \phantom{i} j}^{\phantom{M} i} = L^{Ii}_{\phantom{Ii} k} D_M L_{I \phantom{k} j}^{\phantom{I} k}.
\end{equation}
Demanding that $P$ and $Q$ transform as the corresponding quantities in the ungauged theory~\cite{Bergshoeff:1985mr}, the  following restriction on the structure constants is obtained
\begin{equation}
f_{IJ}^{\phantom{IJ} L} \eta_{KL} = f_{[IJ}^{\phantom{IJ} L}
\eta_{K]L}. \label{e-2-11}
\end{equation}
Any solution of the above equation specifies a consistent gauging
of a subgroup of the isometry group of the scalar manifold. We may
easily find solutions of (\ref{e-2-11}) by taking $\eta_{IJ}$ to be the
Cartan-Killing metric of the gauged algebra. In that case, the
gauge group can either be $SO(3,1)\times H$ with $\dim H =
N\!-\!3$ or $SO(3)\times H$ with ${\rm dim}H=N$. Alternatively, we
may consider $\eta_{IJ}=(\eta_{mn},\delta_{\bar{m} \bar{n}})$,
where $m,n=1,\ldots, p$, $\bar{m},\bar{n}=p+1,\ldots,N+3$ and take
$\eta_{mn}$ to be the Cartan-Killing metric of $SO(3)\times H$,
with $\dim H =p\!-\!3$, or $SO(3,1)\times H$, with ${\rm
dim}H=p\!-\!6$. It should be noted that there are $N+3-p$ $U(1)$
factors of the gauge group in this case.

We define also, for later use, the following projections of the
structure constants $f_{IJ}^{\phantom{IJ} K}$,
\begin{eqnarray}
\label{e-2-12}
C &=& i f_{IJ}^{\phantom{IJ} K} L^{I \phantom{k} i}_{\phantom{I}
k} L^{Jj}_{\phantom{Jj} i} L_{K  \phantom{k} j}^{\phantom{K} k}, \nn\\
C^{ai}_{\phantom{ai} j} &=& i f_{IJ}^{\phantom{IJ} K}
L^{Ii}_{\phantom{Ii} k} L^{Jk}_{\phantom{Jk} j}
L_K^{\phantom{K} a}, \\
C_{ab \phantom{i} j}^{\phantom{ab} i} &=& f_{IJ}^{\phantom{IJ} K}
L_a^{\phantom{a} I} L_b^{\phantom{b} J} L_{K \phantom{i}
j}^{\phantom{K} i}, \nn 
\end{eqnarray}
which can be proven to be the only nonvanishing projections.

\subsection{The 3--form theory}

The Lagrangian for the reducible multiplet ${\cal{V}}_3$ in (\ref{e-2-2}) which contains the
3--form is
{\allowdisplaybreaks \begin{eqnarray}
\label{e-2-13}
E^{-1} \mathcal{L}_7 &=& \frac{1}{2} R - \frac{1}{48} \sigma^{-4}
F_{MNPQ} F^{MNPQ} - \frac{1}{4} \sigma^2 (L_{I \phantom{i}
j}^{\phantom{I} i} L_{J \phantom{j} i}^{\phantom{J} j} +
L_{I}^{\phantom{I} a} L_{Ja} ) F^I_{MN} F^{MNJ}
\nn\\
&&- \frac{1}{2} P_{M \phantom{ai} j}^{\phantom{M} ai} P^{M
\phantom{a} j}_{\phantom{M} a \phantom{j} i} - \frac{1}{2}
\partial_M \phi \partial^M \phi - \frac{1}{2} \bar{\psi}^i_M
\Gamma^{MNP} D_N \psi_{Pi} - \frac{1}{2} \bar{\chi}^i \Gamma^M D_M
\chi_i
\nn\\
&& - \frac{1}{2} \bar{\theta}^{ai} \Gamma^M D_M \theta_{ai}
- \sigma^{-2} \Biggl[ \frac{1}{8 \sqrt{2}} \left(
\bar{\psi}^{Mi} \Gamma^{NP} \psi^Q_i + \frac{1}{12}
\bar{\psi}^i_L \Gamma^{LMNPQR} \psi_{Ri} \right)
\nn\\
&&
- \frac{1}{6
\sqrt{10}} \left( \bar{\chi}^i \Gamma^{MNP}
 \psi^Q_i - \frac{1}{4} \bar{\chi}^i \Gamma^{LMNPQ} \psi_{Li} \right) \nn\\
&& - \frac{1}{160\sqrt{2}} \bar{\chi}^i \Gamma^{MNPQ} \chi_i +
\frac{1}{96 \sqrt{2}} \bar{\theta}^{ai} \Gamma^{MNPQ} \theta_{ai}
\Biggr] F_{MNPQ}
\nn\\
&&- \sigma \Biggl\{ \Biggl[ \frac{i}{2 \sqrt{2}} \left (
\bar{\psi}^{Mi} \psi^N_j + \frac{1}{2} \bar{\psi}^i_L
\Gamma^{LMNP} \psi_{Pj} \right) + \frac{i}{2\sqrt{10}} \left(
\bar{\chi}^i \Gamma^{LMN} \psi_{Lj} \right.
\nn\\ &&
\left.- 2 \bar{\chi}^i \Gamma^M
\psi^N_j \right)
+ \frac{3i}{20\sqrt{2}} \bar{\chi}^i
\Gamma^{MN} \chi_j - \frac{i}{4 \sqrt{2}} \bar{\theta}^{ai}
\Gamma^{MN} \theta^a_j \Biggr] L_{I \phantom{j} i}^{\phantom{I} j}
\nn\\ && +\Biggl[ \frac{1}{4} \left( \bar{\theta}_a^i \Gamma^{LMN}
\psi_{Li} - 2 \bar{\theta}_a^i \Gamma^M \psi^N_i \right) -
\frac{1}{2\sqrt{5}} \bar{\theta}_a^i \Gamma^{MN} \chi_i \Biggr]
L_I^{\phantom{I} a} \Biggr\} F^I_{MN}
\nn\\
&&- \frac{i}{\sqrt{2}} \left( \bar{\theta}^{ai} \psi^M_j -
\bar{\theta}^{ai} \Gamma^{MN} \psi_{Nj} \right) P_{Ma \phantom{j}
i}^{\phantom{Ma} j} + \frac{1}{2} \bar{\chi}^i \Gamma^N \Gamma^M
\psi_{N i} \partial_M \phi
\nn\\
&&+ \frac{i}{\sqrt{2}} \sigma^{-1} \bar{\theta}^{aj} \theta^b_i
C_{ab \phantom{i} j}^{\phantom{ab} i} - \frac{i}{2} \sigma^{-1}
\left( \bar{\psi}^j_M \Gamma^M \theta^a_i + \frac{2}{\sqrt{5}}
\bar{\chi}^j \theta^a_i \right) C_{a \phantom{i} j}^{\phantom{a}
i}
\nn\\
&&- \frac{1}{60 \sqrt{2}} \sigma^{-1} \left( \bar{\psi}^i_M
\Gamma^{MN} \psi_{Ni} + 2\sqrt{5} \bar{\psi}^i_M \Gamma^M \chi_i +
3 \bar{\chi}^i \chi_i - 5 \bar{\theta}^{ai} \theta_{ai} \right) C
\nn\\
&&+ \frac{1}{36} \sigma^{-2} \left( C^2 - 9 C^{ai}_{\phantom{ai}
j}
 C_{a \phantom{j} i}^{\phantom{a} j} \right)
\nn\\
&&+ \frac{1}{48 \sqrt{2}} E^{-1} \epsilon^{MNPQRST} F_{MNPQ}
\Omega_{Y,RST} + (\mathrm{Fermi})^4. 
\end{eqnarray}}
Here, $E^A_{\phantom A M}$ is the siebenbein, $\sigma$ is the
following function of the scalar $\phi$
\begin{equation}
\sigma = \exp \left( - \frac{\phi}{\sqrt{5}} \right),
\label{e-2-14}
\end{equation}
the spinor covariant derivative $D_M$ is defined as
\begin{equation}
D_M \chi_i = \partial_M \chi_i + \frac{1}{4} \omega_{ABM}
\Gamma^{AB} \chi_i + \frac{1}{2} Q_{Mi}^{\phantom{Mi} j} \chi_j
\label{e-2-15}
\end{equation}
and the two field strengths $F_4$ and $F^I_2$ are given by
\begin{equation}
F_{MNPQ} = 4 \partial_{[M} A_{NPQ]} ,\qquad F^I_{MN} = 2
\partial_{[M} A^I_{N]} + f_{JK}^{\phantom{JK} I} A^J_M A^K_N.
\label{e-2-16}
\end{equation}
Also, $\Omega_{Y,MNP}$ is a shorthand for the Chern-Simons form of
the vector multiplets
\begin{equation}
\Omega_{Y,MNP} = \eta_{IJ} F^I_{[MN} A^J_{P]} - \frac{1}{3}
f_{IJ}^{\phantom{IJ} K} A^I_M A^J_N A_{PK}. \label{e-2-17}
\end{equation}

The Lagrangian (\ref{e-2-12}) is invariant under the following set
of local supersymmetry transformations {\allowdisplaybreaks
\begin{eqnarray}
\label{e-2-18}
\delta E^A_{\phantom{A} M} &=& \frac{1}{2}
\bar{\epsilon}^i \Gamma^A \psi_{Mi},
\nn\\
\delta \phi &=& \frac{1}{2} \bar{\epsilon}^i \chi_i,
\nn\\
\delta A_{MNP} &=& \frac{3}{2\sqrt{2}} \sigma^2 \bar{\psi}^i_{[M}
\Gamma_{NP]} \epsilon_i + \frac{1}{\sqrt{10}} \bar{\chi}^i
\Gamma_{MNP} \epsilon_i,
\nn\\
L_{I \phantom{i} j}^{\phantom{I} i} \delta A^I_M &=&
\frac{i}{\sqrt{2}} \sigma^{-1} \left( \bar{\psi}^i_{M} \epsilon_j
- \frac{1}{2} \delta^i_{\phantom{i} j} \bar{\psi}^k_{M} \epsilon_k
\right) - \frac{i}{\sqrt{10}} \sigma^{-1} \left( \bar{\chi}^i
\Gamma_M \epsilon_j - \frac{1}{2} \delta^i_{\phantom{i} j}
\bar{\chi}^k \Gamma_M \epsilon_k \right),
\nn\\
L_{I}^{\phantom{I} a} \delta A^I_M &=& \frac{1}{2} \sigma^{-1}
\bar{\epsilon}^i \Gamma_{M} \theta^a_i,
\nn\\
\delta L_{I \phantom{i} j}^{\phantom{I} i} &=& -
\frac{i}{\sqrt{2}} \left( \bar{\epsilon}^i \theta_{aj} -
\frac{1}{2 N} \delta^i_{\phantom{i} j} \bar{\epsilon}^k
\theta_{ak} \right) L_{I}^{\phantom{I} a},
\\
\delta L_{I}^{\phantom{I} a} &=& - \frac{i}{\sqrt{2}}
\bar{\epsilon}^i \theta^a_j L_{I \phantom{i} i}^{\phantom{I} j},
\nn\\
\delta \psi_{Mi} &=& D_M \epsilon_i + \frac{1}{80 \sqrt{2}}
\sigma^{-2} \left( \Gamma_M^{\phantom{M} NPQR} - \frac{8}{3}
\delta_M^{\phantom{M} N} \Gamma^{PQR} \right) F_{NPQR} \epsilon_i
\nn\\ &&+ \frac{i}{10 \sqrt{2}} \sigma \left(
\Gamma_M^{\phantom{M} NP} - 8 \delta_M^{\phantom{M} N} \Gamma^P
\right) F_{NPi}^{\phantom{NPi}j} \epsilon_i - \frac{1}{30
\sqrt{2}} \sigma^{-1} C\Gamma_M \epsilon_i,
\nn\\
\delta \chi_i &=& \frac{1}{2} \Gamma^M \partial_M \phi \epsilon_i
+ \frac{1}{24 \sqrt{10}} \sigma^{-2} \Gamma^{MNPQ} F_{NPQR}
\epsilon_i \nn \\ &&- \frac{i}{2 \sqrt{10}} \sigma \Gamma^{MN}
F_{MNi}^{\phantom{MNi}j} \epsilon_j + \frac{1}{6 \sqrt{10}}
\sigma^{-1} C \epsilon_i,
\nn\\
\delta \theta^a_i &=& - \frac{1}{4} \sigma \Gamma^{MN} F^I_{MN}
L_I^{\phantom{I} a} \epsilon_i + \frac{1}{\sqrt{2}} \Gamma^M P_{M
\phantom{a} i}^{\phantom{M} a \phantom{i} j} \epsilon_j -
\frac{i}{2} \sigma^{-1} C^{aj}_{\phantom{aj} i} \epsilon_j \nn.
\end{eqnarray}}
where $\epsilon_i$ is a symplectic Majorana spinor.

We should here remark that the action of the theory may in
principle include the topological mass term
\begin{equation}
S_{m} = \frac{h}{36} \int d^7x \epsilon^{MNPQRST} F_{MNPQ}
A_{RST}, \label{e-2-19}
\end{equation}
which is present in the pure supergravity case (no vector
multiplets). This term has important implications for the
6-dimensional theory obtained by reduction on ${\bf S}^1$
\cite{Giani:dw} while, in the context of ${\bf S}^1/\mathbb{Z}_2$
compactification, its variation contributes to the anomaly and the
anomaly cancellation conditions lead to a relation fixing the
boundary Yang-Mills coupling in terms of the gravitational
coupling and $h$~\cite{Gherghetta:2002xf,Gherghetta:2002nq}. In
addition, this term explicitly depends on $A_3$ and thus it is not
possible to perform a duality transformation to obtain an
equivalent theory with a 2--form potential. However, in the
presence of vector multiplets, this term should vanish $(h=0)$ by
supersymmetry and thus, it does not exist for the 7D ${\cal{N}}=2$
supergravity coupled to vector multiplets~\cite{Park:id}.

\subsection{Duality transformation in the presence of boundaries}

It is well-known that the ambiguity in the representation of
antisymmetric tensor fields, emanating from the fact that a
$p$--form potential and a $(D-p-2)$--form potential contain
exactly the same degrees of freedom, often allows us to express a
given theory involving such a field in two dual formulations. In
such a case, the two dual theories can be obtained one from the
other by means of a duality transformation
\cite{Cremmer:1979up,Nicolai:1980td}.

In the context of $D=7$, $\mathcal{N}=2$ supergravity, there exist
two such formulations, so that the 3--form theory considered above
has a dual formulation in terms of a 2--form potential. The
equivalence of the two theories under a duality transformation was
demonstrated in \cite{Park:id}, where the 2--form theory was
obtained by dualizing the 3--form one. Here, we shall repeat the
same construction, this time on a manifold with boundary. The
difference in this case is that the duality transformation yields
an extra boundary term that will play a particular role in the
boundary theory.

The first step in the duality transformation of the 3--form
potential $A_3$ is to consider the terms of (\ref{e-2-12}) and
(\ref{e-2-17}) that involve the field strength $F_4$ and replace
the latter by a new unconstrained field $S_4$ whose SUSY
transformation is taken to be the same as that for $F_4$. The
modified Lagrangian may be written as
\begin{eqnarray}
\label{e-2-20}
\mathcal{L}_S &=& - \frac{1}{48} E \sigma^{-4} S_{MNPQ} S^{MNPQ} -
\frac{1}{8 \sqrt{2}} E \sigma^{-2} S_{MNPQ} J^{MNPQ} \nn \\ && +
\frac{1}{48 \sqrt{2}} \epsilon^{MNPQRST} S_{MNPQ} \Omega_{Y,RST} +
\ldots 
\end{eqnarray}
where the dots stand for the rest of the terms in (\ref{e-2-13})
and where we defined
\begin{eqnarray}
\label{e-2-21}
J_{MNPQ} &=& \bar{\psi}^i_{[M} \Gamma_{NP} \psi_{Q]i} +
\frac{1}{12} \bar{\psi}^{Li} \Gamma_{LMNPQR} \psi^R_i - \frac{4}{3
\sqrt{5}} \biggl( \bar{\chi}^i \Gamma_{[MNP} \psi_{Q]i} \nn
\\ && - \frac{1}{4} \bar{\chi}^i \Gamma_{LMNPQ} \psi^L_i
\biggr) - \frac{1}{20} \bar{\chi}^i \Gamma_{MNPQ} \chi_i +
\frac{1}{12} \bar{\theta}^{ai}
 \Gamma_{MNPQ} \theta_{ai}.
\end{eqnarray}

The next step is to add the Lagrange-multiplier term
\begin{equation}
\mathcal{L}_C = - \frac{1}{48} \epsilon^{MNPQRST} B_{MN}
\partial_P S_{QRST}, \label{e-2-22}
\end{equation}
where $B_2$ is a new unconstrained field. Varying $\mathcal{L}_C$
with respect to $B_2$ would simply enforce the Bianchi identity $d
S_4 = 0$, taking us back to the original theory. Since in that
case, $\mathcal{L}_S$ would be SUSY-invariant, the SUSY variation
of $\mathcal{L}_S$ for arbitrary $S_4$ is a term proportional to
$\partial_{[M} S_{NPQR]}$ arising from the variations of the
fermion kinetic terms. Writing this term as
\begin{equation}
\delta \mathcal{L}_S = \frac{1}{48} \epsilon^{MNPQRST} X_{MN}
\partial_P S_{QRST}, \label{e-2-23}
\end{equation}
and taking $B_2$ to transform according to
\begin{equation}
\delta B_{MN} = X_{MN}, \label{e-2-24}
\end{equation}
we see that the ``intermediate'' Lagrangian,
\begin{equation}
\mathcal{L}_I \equiv \mathcal{L}_S + \mathcal{L}_C, \label{e-2-25}
\end{equation}
is locally supersymmetric since
\begin{eqnarray}
\label{e-2-26}
\delta \mathcal{L}_I &= &- \frac{1}{48} \epsilon^{MNPQRST} B_{MN}
\partial_P \delta S_{QRST} \nn \\
&=& - \frac{1}{12} \epsilon^{MNPQRST} B_{MN} \partial_P \partial_Q
\delta A_{RST} = 0.
\end{eqnarray}
We also note here that if $F_4$ were to satisfy, instead of $d F_4
= 0$, a modified Bianchi identity of the type $d F_4 = I_5$ for
some 5--form $I_5$, (\ref{e-2-22}) would have to be modified to
\begin{equation}
\mathcal{L}_C = - \frac{1}{48} \epsilon^{MNPQRST} B_{MN} \left(
\partial_{P} S_{QRST} - \frac{1}{5} I_{PQRST} \right),
\label{e-2-26a}
\end{equation}
so as to enforce this Bianchi identity on $S_4$ when varied with
respect to $B_2$.

The next step is to integrate the Lagrange multiplier term
(\ref{e-2-22}) by parts. After that, $\mathcal{L}_I$ can be
written in the form
\begin{equation}
\mathcal{L}_I = \mathcal{L}_{I,bulk} + \mathcal{L}_{I,bdy},
\label{e-2-27}
\end{equation}
where $\mathcal{L}_{I,bulk}$ contains the bulk terms
\begin{eqnarray}
\label{e-2-28}
\mathcal{L}_{I,bulk} &=& - \frac{1}{48} E \sigma^{-4} S_{MNPQ}
S^{MNPQ} - \frac{1}{8 \sqrt{2}} E \sigma^{-2} J^{MNPQ} S_{MNPQ}
\nn
\\ &&
 + \frac{1}{144} \epsilon^{MNPQRST} S_{MNPQ} G_{RST} + \ldots
\end{eqnarray}
and $\mathcal{L}_{I,bdy}$ is the surface term
\begin{equation}
\mathcal{L}_{I,bdy} = - \frac{1}{48} \epsilon^{MNPQRST}
\partial_M (B_{NP} S_{QRST}), \label{e-2-29}
\end{equation}
which, in the presence of boundaries, does not a priori vanish. In
(\ref{e-2-28}), $G_3$ stands for the (modified) field strength of
$B_2$, defined by
\begin{equation}
G_{MNP} = 3 \left[ \partial_{[M} B_{NP]} + \frac{1}{\sqrt{2}}
\left( \eta_{IJ} F^I_{[MN} A^J_{P]} - \frac{1}{3}
f_{IJ}^{\phantom{IJ} K} A^I_M A^J_N A_{PK} \right) \right].
\label{e-2-30}
\end{equation}
The intermediate SUSY transformation laws for $\psi_{Mi}$ and
$\chi_i$ are given by those in (\ref{e-2-18}) with $F_4$ replaced
by $S_4$. Hence, we have
\begin{equation}
\delta \psi_{Mi} = \frac{1}{80 \sqrt{2}} \sigma^{-2} \left(
\Gamma_M^{\phantom{M} NPQR} - \frac{8}{3} \delta_M^{\phantom{M} N}
\Gamma^{PQR} \right) S_{NPQR} \epsilon_i + \ldots \label{e-2-31}
\end{equation}
and
\begin{equation}
\delta \chi_i = \frac{1}{24 \sqrt{10}} \sigma^{-2} \Gamma^{MNPQ}
S_{MNPQ} \epsilon_i + \ldots \label{e-2-32}
\end{equation}
where the dots correspond to the rest of the terms in
(\ref{e-2-18}).

The final step in the duality transformation is to integrate out
$S_4$. On the bulk, this can be accomplished by using its
(algebraic) equation of motion
\begin{equation}
S^{MNPQ} = \frac{1}{6} E^{-1} \sigma^4 \epsilon^{MNPQRST} G_{RST}
- \frac{3}{\sqrt{2}} \sigma^2 J^{MNPQ} \label{e-2-33}
\end{equation}
Substituting this result in the Lagrangian (\ref{e-2-13}) and the
SUSY transformation rules (\ref{e-2-18}), one obtains a dual
theory, phrased in terms of the 2--form potential $B_2$, whose
Lagrangian and transformation rules will be stated shortly. In the
absence of spacetime boundaries, this theory would be equivalent
to the 2--form version of $D=7$, $\mathcal{N}=2$ supergravity
whose special cases were constructed in
\cite{Salam:1983fa,Bergshoeff:1985mr}. However, in the presence of
a boundary, it is not: the former theory contains the surface term
(\ref{e-2-29}) not present in the latter. Although one could think
of invoking the Bianchi identity of $F_4$ to set the value of this
term to zero, this is not correct: in the 3--form version, anomaly
and supersymmetry considerations result in a modified Bianchi
identity for $F_4$, as in the HW case. This in turn induces a
nonvanishing boundary value for $S_4$ so that the surface term
gives rise to a 6D boundary interaction.

\subsection{The 2--form theory}

The gauged 2--form version of $D=7$, $\mathcal{N}=2$ supergravity obtained by the duality transformation discussed above constitutes a generalization of the theory constructed in \cite{Bergshoeff:1985mr} for any subgroup of the holonomy group. The resulting Lagrangian for the multiplet ${\cal{V}}_2$ in (\ref{e-2-3}) is given by {\allowdisplaybreaks
\begin{eqnarray}
\label{e-2-34}
E^{-1} \mathcal{L}_7 &=&
\frac{1}{2} R - \frac{1}{12} \sigma^4 G_{MNP} G^{MNP} -
\frac{1}{4} \sigma^2 (L_{I \phantom{i} j}^{\phantom{I} i} L_{J
\phantom{j} i}^{\phantom{J} j} + L_{I}^{\phantom{I} a} L_{Ja} )
F^I_{MN} F^{MNJ}
\nn\\
&&- \frac{1}{2} P_{M \phantom{ai} j}^{\phantom{M} ai} P^{M
\phantom{a} j}_{\phantom{M} a \phantom{j} i}  - \frac{1}{2}
\partial_M \phi \partial^M \phi - \frac{1}{2} \bar{\psi}^i_M
\Gamma^{MNP} D_N \psi_{Pi}
\nn \\
&&
- \frac{1}{2} \bar{\chi}^i \Gamma^M D_M
\chi_i - \frac{1}{2} \bar{\theta}^{ai} \Gamma^M D_M \theta_{ai}
+ \frac{1}{48 \sqrt{2}} \sigma^2 \Biggl[ - 2 \bar{\psi}^{Li}
\Gamma_{[L} \Gamma^{MNP} \Gamma_{Q]} \psi^Q_i
\nn\\
&&
+ \frac{8}{\sqrt{5}}
\bar{\chi}^i \Gamma^L \Gamma^{MNP} \psi_{Li} + \frac{6}{5}
\bar{\chi}^i \Gamma^{MNP} \chi_i - 2 \bar{\theta}^{ai}
\Gamma^{MNP} \theta_{ai} \Biggr] G_{MNP}
\nn\\
&&- \sigma \Biggl\{ \Biggl[ \frac{i}{2 \sqrt{2}} \left (
\bar{\psi}^{Mi} \psi^N_j + \frac{1}{2} \bar{\psi}^i_L
\Gamma^{LMNP} \psi_{Pj} \right) + \frac{i}{2\sqrt{10}}
\left( \bar{\chi}^i \Gamma^{LMN} \psi_{Lj} \right.
\nn\\
&&
\left.- 2 \bar{\chi}^i \Gamma^M \psi^N_j \right)
 + \frac{3i}{20\sqrt{2}} \bar{\chi}^i \Gamma^{MN} \chi_j - \frac{i}{4 \sqrt{2}}
\bar{\theta}^{ai} \Gamma^{MN} \theta_{a j} \Biggr] L_{I \phantom{j} i}^{\phantom{I} j} \nn\\
&& +\Biggl[ \frac{1}{4} \left( \bar{\theta}_a^i \Gamma^{LMN}
\psi_{Li} - 2 \bar{\theta}_a^i \Gamma^M \psi^N_i \right) -
\frac{1}{2\sqrt{5}} \bar{\theta}_a^i \Gamma^{MN} \chi_i \Biggr]
L_I^{\phantom{I} a} \Biggr\} F^I_{MN}
\nn\\
&&- \frac{i}{\sqrt{2}} \left( \bar{\theta}^{ai} \psi^M_j -
\bar{\theta}^{ai} \Gamma^{MN} \psi_{Nj} \right) P_{Ma \phantom{j}
i}^{\phantom{Ma} j} + \frac{1}{2} \bar{\chi}^i \Gamma^N \Gamma^M
\psi_{N i} \partial_M \phi
\nn\\
&&+ \frac{i}{\sqrt{2}} \sigma^{-1} \bar{\theta}^{aj} \theta^b_i
C_{ab \phantom{i} j}^{\phantom{ab} i} - \frac{i}{2} \sigma^{-1}
\left( \bar{\psi}^j_M \Gamma^M \theta^a_i + \frac{2}{\sqrt{5}}
\bar{\chi}^j \theta^a_i \right) C_{a \phantom{i} j}^{\phantom{a}
i}
\nn\\
&&- \frac{1}{60 \sqrt{2}} \sigma^{-1} \left( \bar{\psi}^i_M
\Gamma^{MN} \psi_{Ni} + 2\sqrt{5} \bar{\psi}^i_M \Gamma^M \chi_i +
3 \bar{\chi}^i \chi_i - 5 \bar{\theta}^{ai} \theta_{ai} \right) C
\nn\\
&&+ \frac{1}{36} \sigma^{-2} \left( C^2 - 9 C^{ai}_{\phantom{ai}
j} C_{a \phantom{j} i}^{\phantom{a} j} \right) + (\mathrm{Fermi})^4. 
\end{eqnarray}}We note that the $(\mathrm{Fermi})^4$ terms are not those in
(\ref{e-2-18}), but receive an additional contribution from a $J_4
\wedge * J_4$ term arising from the second term of (\ref{e-2-33}).

The SUSY transformation rules for this theory are given by
{\allowdisplaybreaks
\begin{eqnarray}
\label{e-2-35}
\delta E^A_{\phantom{A} M} &=& \frac{1}{2}
\bar{\epsilon}^i \Gamma^A \psi_{Mi},
\nn\\
\delta \phi &=& \frac{1}{2} \bar{\epsilon}^i \chi_i,
\nn\\
\delta B_{MN} &=& \sigma^2 \left( - \frac{1}{\sqrt{2}}
\bar{\epsilon}^i \Gamma_{[M} \psi_{N] i} - \frac{1}{\sqrt{5}}
\bar{\epsilon}^i \Gamma_{MN} \chi_i \right) + \frac{1}{\sqrt{2}}
A^I_{[M} \delta A^J_{N]} \eta_{IJ},
\nn\\
L_{I \phantom{i} j}^{\phantom{I} i} \delta A^I_M &=&
\frac{i}{\sqrt{2}} \sigma^{-1} \left( \bar{\psi}^i_{M} \epsilon_j
- \frac{1}{2} \delta^i_{\phantom{i} j} \bar{\psi}^k_{M} \epsilon_k
\right) - \frac{i}{\sqrt{10}} \sigma^{-1} \left( \bar{\chi}^i
\Gamma_M \epsilon_j - \frac{1}{2} \delta^i_{\phantom{i} j}
\bar{\chi}^k \Gamma_M \epsilon_k \right),
\nn\\
L_{I}^{\phantom{I} a} \delta A^I_M &=& \frac{1}{2} \sigma^{-1}
\bar{\epsilon}^i \Gamma_{M} \theta^a_i,
\nn\\
\delta L_{I \phantom{i} j}^{\phantom{I} i} &=& -
\frac{i}{\sqrt{2}} \left( \bar{\epsilon}^i \theta_{aj} -
\frac{1}{2 N} \delta^i_{\phantom{i} j} \bar{\epsilon}^k
\theta_{ak} \right) L_{I}^{\phantom{I} a},
\\
\delta L_{I}^{\phantom{I} a} &=& - \frac{i}{\sqrt{2}}
\bar{\epsilon}^i \theta^a_j L_{I \phantom{i} i}^{\phantom{I} j},
\nn\\
\delta \psi_{Mi} &=& D_M \epsilon_i - \frac{1}{120 \sqrt{2}}
\sigma^{2} \left( \Gamma_M \Gamma^{NPQ} + 5 \Gamma^{NPQ} \Gamma_M
\right) G_{NPQ} \epsilon_i \nn\\ &&+ \frac{i}{10 \sqrt{2}} \sigma
\left( \Gamma_M^{\phantom{M} NP} - 8 \delta_M^{\phantom{M} N}
\Gamma^P \right) F_{NPi}^{\phantom{NPi}j} \epsilon_i - \frac{1}{30
\sqrt{2}} \sigma^{-1} C\Gamma_M \epsilon_i,
\nn\\
\delta \chi_i &=& \frac{1}{2} \Gamma^M \partial_M \phi \epsilon_i
- \frac{1}{6 \sqrt{10}} \sigma^{2} \Gamma^{MNP} G_{MNP} \epsilon_i
\nn \\ && - \frac{i}{2 \sqrt{10}} \sigma \Gamma^{MN}
F_{MNi}^{\phantom{MNi}j} \epsilon_j + \frac{1}{6 \sqrt{10}}
\sigma^{-1} C \epsilon_i,
\nn\\
\delta \theta^a_i &=& - \frac{1}{4} \sigma \Gamma^{MN} F^I_{MN}
L_I^{\phantom{I} a} \epsilon_i + \frac{1}{\sqrt{2}} \Gamma^M P_{M
\phantom{a} i}^{\phantom{M} a \phantom{i} j} \epsilon_j -
\frac{i}{2} \sigma^{-1} C^{aj}_{\phantom{aj} i} \epsilon_j \nn.
\end{eqnarray}}

\section{Orbifold compactification}
\label{s-3}

The potential of the $3N+1$ scalars of the theory (\ref{e-2-34})
is given by
\begin{equation}
V(\phi^\alpha,\phi)=\frac{1}{36} \sigma^{-2} \left( 9
C^{ai}_{\phantom{ai} j} C_{a \phantom{j} i}^{\phantom{a} j}-C^2
\right) \label{pot}
\end{equation}
where the projections $C$ and $C^{ai}_{\phantom{ai} j}$,
introduced in (\ref{e-2-12}), are functions of the scalars
$\phi^\alpha$ and $\phi$ and depend on the structure constants of
the gauge group. Although a general result for the critical points
of the potential (\ref{pot}) does not exist, it is not difficult
to see that the theory possesses a seven-dimensional Minkowski
vacuum. Indeed, we may consider the following scalar configuration
\begin{eqnarray}
\label{LL}
L_{\bar{n}}^{\phantom{\bar{n}} n} = \delta_{\bar{n}}^n ~~~
\mbox{for}~~n=1,2,3 ~~~
\mbox{and}~~~L_{\bar{a}}^{\phantom{\bar{a}} a} =\delta_{\bar{a}}^a~~~\mbox{for}~~a=1,...,N,
\end{eqnarray}
where we have split the index $I=1,....,N+3$ as $I=(\bar{n},\bar{a})$\footnote{In the doublet notation, the first of (\ref{LL}) translates to $L_{\bar{n} \phantom{i} j}^{\phantom{\bar{n}} i} = \frac{1}{\sqrt{2}} (\sigma_{\bar{n}})^i_{\phantom{i} j}$.}. It is not hard to see that (\ref{LL}) satisfies (\ref{ij}) and that, for an appropriate choice of the gauge group, we have
\begin{equation}
C={C_a}^i_j={C_{ab}}^i_j=0.
\label{CC}
\end{equation}
Then, the potential and its derivatives with respect to $\phi^a$
vanish for the choice (\ref{LL}) so that $\phi^a,\phi={\rm
const.}$ is a solution, which gives rise to a 7D Minkowski vacuum.

The theory can be dimensionally reduced to six dimensions along an
${\bf S}^1$ of radius $R$ parameterized by $x_7$. On ${\bf S}^1$,
the various 7-dimensional fields of the reducible multiplet
(\ref{e-2-3}) decompose \`a la Kaluza-Klein according to
\begin{eqnarray}
\label{e-3-1}
&g_{MN} \to g_{\mu\nu} , A_\mu , \widetilde{\xi} \quad,\quad
 B_{MN} \to B_{\mu\nu},B_\mu \quad,\quad A_M^I
\to A_\mu^I , A^I \quad,\quad \phi^\alpha,\phi \to \phi^\alpha,
\widetilde{\phi} \nn \\ &\psi^i_M \to \psi^i_\mu,
\widetilde{\psi}^i \quad,\quad \chi^i \to \widetilde{\chi}^i
\quad, \quad \theta^{ai} \to \theta^{ai}. 
\end{eqnarray}
The detailed reduction procedure was presented in \cite{Giani:dw}.
In the reduction presented therein, the Kaluza-Klein ansatz for
the 7D metric reads
\begin{equation}
d s_7^2 = e^{ - \widetilde{\xi} / \sqrt{5} } d s_6^2 + e^{
4\widetilde{\xi} / \sqrt{5} } \left( d x_7 + A_\mu dx^\mu
\right)^2, \label{e-3-2}
\end{equation}
the various 7D spinors reduce according to
\begin{equation}
\psi^{(7)}_{\mu i} = e^{ - \widetilde{\xi} / 4 \sqrt{5} } \left[
\psi_{\mu i} - \frac{1}{2 \sqrt{5}} \left( \Gamma_\mu \Gamma_7 + 4
\sqrt{2} A_\mu \right) \widetilde{\psi}_i \right], \label{e-3-3}
\end{equation}
\begin{equation}
\psi^{(7)}_{7 i} = \frac{2}{\sqrt{5}} e^{ \widetilde{\xi} /
4\sqrt{5} } \widetilde{\psi}_i \quad,\quad\chi^{(7)}_{i} = e^{
\widetilde{\xi} / 4\sqrt{5} } \widetilde{\chi}_i \quad,\quad
\theta^{(7)}_{ai} = e^{ \widetilde{\xi} / 4\sqrt{5} }
\theta_{ai}, \label{e-3-4}
\end{equation}
while antisymmetric tensor fields reduce in the usual way. In the
above, $\psi^i_\mu$ is identified with the 6D gravitino. As for
the tilded 6D fields $( \widetilde{\psi}_i , \widetilde{\chi}_i )$
and $( \widetilde{\xi} , \widetilde{\phi} )$, it helps to trade
them for the fields $( \psi_i , \chi_i )$ and $( \xi , \phi )$
defined through the linear combinations
\begin{equation}
\widetilde{\psi}_i = \frac{1}{\sqrt{5}} \left( 2 \psi_i - \Gamma_7
\chi_i \right) ~,\qquad \widetilde{\chi}_i = \frac{1}{\sqrt{5}}
\left( 2 \chi_i + \Gamma_7 \psi_i \right), \label{e-3-5}
\end{equation}
and
\begin{equation}
\widetilde{\xi} = \frac{1}{\sqrt{5}} \left( 2 \xi - \phi \right)
~,\qquad \widetilde{\phi} = \frac{1}{\sqrt{5}} \left( 2 \phi
+ \xi \right). \label{e-3-6}
\end{equation}
The Lagrangian of the theory on ${\bf S}^1$ is then obtained by
substituting (\ref{e-3-2}--\ref{e-3-6}) into (\ref{e-2-34}). The
final result is a $D=6$, $\mathcal{N}=4$ supergravity model, which
can be consistently truncated to a model with $\mathcal{N}=2$
supersymmetry.

In what follows, we will consider compactification of our theory on the ${\bf S}^1/\mathbb{Z}_2$ orbifold, in order to obtain a model with $\mathcal{N}=1$ supersymmetry. In contrast to common practice in bulk-brane theories, we will make the somewhat unconventional choice of writing the 7D bulk fields in the ``Kaluza-Klein basis'' described by the preceding equations. This particular choice is dictated, in our case, by the fact that we intend to explore in detail the structure of the boundary theory, which is most conveniently done in a basis adapted for 6D fields. To avoid notational confusion, subsequent references to bulk fields will be always accompanied by a ``$(7)$'' superscript.

\subsection{Compactification on ${\bf S}^1 / \mathbb{Z}_2$}

To obtain a chiral 6-dimensional theory from our model, we
consider compactification of the $x_7$ coordinate on the ${\bf
S}^1/\mathbb{Z}_2$ orbifold. The $\mathbb{Z}_2$ action is as usual
$x_7\to -x_7$ and the two fixed points are at $x_7=0$ and $x_7=\pi
R$. We should note that we may consider  two different, but
equivalent, approaches. In the ``downstairs'' approach, one
considers the 7D spacetime manifold as the product $M_6 \times I$
of 6D spacetime times the interval $I=[0,\pi R]$ obtained by
modding out the $\mathbb{Z}_2$ symmetry. Since this is a
\emph{manifold with boundary}, possible surface terms resulting
from partial integrations need to be retained. In the ``upstairs''
approach, which is the one originally used by Ho\v rava and
Witten, one regards spacetime as the smooth manifold $M_6 \times
{\bf S}^1$ subject to $\mathbb{Z}_2$ invariance. On this
\emph{orbifold}, no surface terms arise from partial integration;
however, the fact that $\mathbb{Z}_2$ has fixed points requires
that certain $\mathbb{Z}_2$-odd functions have step-function
discontinuities. In what follows, we shall mainly work in the
``downstairs'' approach.

To begin the construction of the theory, we have to mod out ${\bf
S}^1$ by $\mathbb{Z}_2$. In order for the reduction to be
consistent, we must first make certain parity assignments to the
fields so that the Lagrangian and the supersymmetry transformation
rules stay invariant under $x_7 \to -x_7$. Then, when we mod out
$\mathbb{Z}_2$, only fields of even parity survive on the two
orbifold fixed planes. By the same considerations as
in~\cite{Bagger:2002rw,Gherghetta:2002xf,Gherghetta:2002nq}, it is
easy to see that, after the $\mathbb{Z}_2$ projection, the
surviving bosonic fields are
\begin{equation}
g_{\mu\nu} , B_{\mu\nu} , A^I , \phi^\alpha , \phi , \xi,
\label{e-3-7}
\end{equation}
while the spinors are subject to the chirality constraints
\begin{equation}
\Gamma_7 \left( \psi^i_\mu , \epsilon^i \right) = - \left(
\psi^i_\mu , \epsilon^i \right) ~,\qquad \Gamma_7 \left(
\psi^i, \chi^i , \theta^{ai} \right) = \left( \psi^i, \chi^i ,
\theta^{ai} \right). \label{e-3-8}
\end{equation}
The surviving fields can be then arranged into multiplets that
furnish representations of the $\mathcal{N}=(0,1)$, $D=6$
supersymmetry algebra. To perform the decomposition, we split
$B_{\mu\nu}$ into a self-dual and an anti-self-dual part
\begin{equation}
B_{\mu\nu} = B^+_{\mu\nu} + B^-_{\mu\nu},
\label{e-3-9}
\end{equation}
and we group all scalars except from $\phi$ and all spin 1/2
fermions except from $\chi^{i+}$ according to
\begin{equation}
\Phi^{Yz} = (A^I,\phi^\alpha,\xi) ~,\qquad Z^{Yi+} =
(\psi^{i+},\theta^{ai+}), \label{e-3-10}
\end{equation}
where the index $Y$ runs from $1$ to $N+1$ while $z$ runs from $1$
to $4$. Then, the fields surviving the $\mathbb{Z}_2$ projection
can be arranged as follows
\begin{eqnarray}
\label{e-3-11}
\text{Gravity multiplet} \quad&:&\quad ( g_{\mu\nu} , B^+_{\mu\nu} , \psi^{i-}_\mu ),\nn\\
\text{Tensor multiplet} \quad&:&\quad ( B^-_{\mu\nu} , \phi , \chi^{i+} ), \\
N+1 \text{ hypermultiplets} \quad&:&\quad (\Phi^{Yz}, Z^{Yi+}
)\nn. 
\end{eqnarray}
where the $\pm$ superscripts on spinors indicate six-dimensional
chirality. Thus, the massless spectrum of the ${\bf S}^1 /
\mathbb{Z}_2$ compactification consists of the graviton multiplet,
a tensor and $N+1$ hypermultiplets.

In the rest of this paper, we shall study the theory pertaining to
a single fixed plane which we may take to be $x_7=0$. Appropriate
modifications that apply to the other fixed plane will be
indicated when necessary.

\subsection{The dimensionally reduced bulk Lagrangian}

The Lagrangian and the supersymmetry transformation rules of the
$D=6$ theory arising from the dimensional reduction of the bulk
theory can be obtained in a straightforward way by substituting
(\ref{e-3-2}--\ref{e-3-6}) (with $\Gamma_7 = 1$ and $A_\mu=0$) in
(\ref{e-2-34}) and (\ref{e-2-35}), neglecting all
$\mathbb{Z}_2$--odd fields and taking account of the chirality
constraints satisfied by the spinors. To keep things relatively
simple, we will neglect the terms involving the bulk
hypermultiplets (their treatment is similar to the other bulk
scalars and fermions), although we will discuss some of their
consequences in what follows.

Let us start from the Lagrangian. By the reduction rules, it is
clear that all kinetic terms retain their canonical form, apart
from a different scalar function in front of $G_3 \wedge * G_3$.
The only interactions  that survive when we ignore the
hypermultiplets are $\bar{\chi} \Gamma \psi \partial \phi$,
$\bar{\psi} \Gamma \psi G$, $\bar{\chi} \Gamma \psi G$ and
$\bar{\chi} \Gamma \chi G$. The former two retain the same form as
in the original action, while the latter two reduce to
\begin{eqnarray}
\label{e-3-12}
&& \frac{1}{48 \sqrt{2}} e^\phi \left( - \frac{4}{\sqrt{5}}
\bar{\psi}_\lambda^i \Gamma^\lambda \Gamma^{\mu\nu\rho}
\widetilde{\psi}_i + \frac{8}{\sqrt{5}} \bar{\widetilde{\chi}}^i
\Gamma^\lambda \Gamma^{\mu\nu\rho} \psi_{\lambda i}
\right) G_{\mu\nu\rho} \nn\\
&\to&  \frac{1}{48 \sqrt{2}} e^\phi \left( \frac{4}{5} +
\frac{16}{5} \right) \bar{\chi}^i \Gamma^\lambda
\Gamma^{\mu\nu\rho} \psi_{\lambda i} G_{\mu\nu\rho} = \frac{1}{12
\sqrt{2}} \bar{\chi}^i \Gamma^\lambda \Gamma^{\mu\nu\rho}
\psi_{\lambda i} G_{\mu\nu\rho}
\end{eqnarray}
and
\begin{eqnarray}
\label{e-3-13}
&& \frac{1}{48 \sqrt{2}} e^\phi \left( - \frac{1}{5}
\bar{\widetilde{\psi}}^{i} \Gamma^{\mu\nu\rho} \Gamma_{\sigma}
\Gamma^\sigma \widetilde{\psi}_i - \frac{16}{5}
\bar{\widetilde{\chi}}^i \Gamma^{\mu\nu\rho} \widetilde{\psi}_{i}
+ \frac{6}{5} \bar{\widetilde{\chi}}^i
\Gamma^{\mu\nu\rho} \widetilde{\chi}_i \right) G_{\mu\nu\rho} \nn\\
&\to& \frac{1}{48 \sqrt{2}} e^\phi \left( - \frac{6}{25} +
\frac{32}{25} + \frac{24}{25}\right) \bar{\chi}^{i}
\Gamma^{\mu\nu\rho} \chi_i  G_{\mu\nu\rho} = \frac{1}{24 \sqrt{2}}
e^\phi \bar{\chi}^{i} \Gamma^{\mu\nu\rho} \chi_i  G_{\mu\nu\rho}
\end{eqnarray}
Therefore, the dimensionally reduced Lagrangian reads
\begin{eqnarray}
\label{e-3-14}
e^{-1} \mathcal{L}_{bulk} &=& \frac{1}{2} R - \frac{1}{12} e^{2
\phi} G_{\mu\nu\rho} G^{\mu\nu\rho} - \frac{1}{2} \partial_\mu
\phi \partial^\mu \phi - \frac{1}{2} \bar{\psi}^i_\mu
\Gamma^{\mu\nu\rho} D_\nu \psi_{\rho i} - \frac{1}{2} \bar{\chi}^i
\Gamma^\mu D_\mu \chi_i
\nn\\
&&- \frac{1}{24 \sqrt{2}} e^\phi \left( \bar{\psi}^{\lambda i}
\Gamma_{[\lambda} \Gamma^{\mu\nu\rho} \Gamma_{\sigma]}
\bar{\psi}^\sigma_i- 2 \bar{\chi}^i \Gamma^\lambda
\Gamma^{\mu\nu\rho} \psi_{\lambda i} - \bar{\chi}^{i}
\Gamma^{\mu\nu\rho} \chi_i \right) G_{\mu\nu\rho}
\nn\\
&&+ \frac{1}{2} \bar{\chi}^i \Gamma^\nu \Gamma^\mu \psi_{\nu i}
\partial_\mu \phi + (\mathrm{Fermi})^4. 
\end{eqnarray}

Passing on to the supersymmetry transformation laws, one has again
to substitute (\ref{e-3-2}--\ref{e-3-6}) in the transformation
laws of the 7D theory. We easily find{\allowdisplaybreaks
\begin{eqnarray}
\label{e-3-15}
\delta e^a_{\phantom{A} \mu} &=& \frac{1}{2} \bar{\epsilon}^i \Gamma^a \psi_{\mu i},
\nn\\
\delta \phi &=& \frac{1}{2} \bar{\epsilon}^i \chi_i,
\nn\\
\delta B_{\mu\nu} &=& e^{-\phi} \left( - \frac{1}{\sqrt{2}}
\bar{\epsilon}^i \Gamma_{[\mu} \psi_{\nu] i} + \frac{1}{2
\sqrt{2}} \bar{\epsilon}^i \Gamma_{\mu\nu} \chi_i \right), \nn\\
\delta \psi_{\mu i} &=& D_\mu \epsilon_i - \frac{1}{24 \sqrt{2}}
e^\phi \Gamma^{\nu\rho\sigma} \Gamma_\mu G_{\nu\rho\sigma}
\epsilon_i, \\
\delta \chi_i &=& \frac{1}{2} \Gamma^\mu
\partial_\mu \phi \epsilon_i - \frac{1}{12 \sqrt{2}} e^\phi
\Gamma^{\mu\nu\rho} G_{\mu\nu\rho} \epsilon_i.\nn 
\end{eqnarray}}
The Lagrangian (\ref{e-3-14}) and the transformation rules
(\ref{e-3-15}) are exactly the same as those appearing in the
fully truncated theory presented in \cite{Giani:dw}.

Although we have not explicitly included the contributions of the
bulk hypermultiplets in our discussion, there is a particular
effect that has to be discussed. In particular, it turns out that
the combination $\sigma^{-1} C$ appearing in the supersymmetry
variation of $\psi_M^i$ in (\ref{e-2-35}) is $\mathbb{Z}_2$--odd
and thus gives rise to delta-function terms on the orbifold fixed
points. On the $x_7=0,\pi R$ fixed planes, these can be cancelled
if we introduce the boundary action
\begin{equation}
S_0 = - \frac{\sqrt{2}}{3} \int d^7 x \, E\,  \sigma^{-1}\,  C\,
\Big{(}\delta(x_7)-\delta(x^7-\pi R)\Big{)}. \label{e-3-16}
\end{equation}
It is easy to see then that the boundary value of
$C=C(\phi^\alpha)$ and $\phi$ specifies $S_0$, which is nothing
else that the tension of the branes at $x^7=0,\pi R$.

\section{Boundary multiplets}
\label{sec-4}

As we have seen in the previous section, the bulk fields surviving
the orbifold projection arrange themselves into a gravity
multiplet, a tensor multiplet and $N+1$ hypermultiplets of the
$D=6$, $\mathcal{N} = 1$ supersymmetry algebra. Since the above
spectrum is chiral the theory suffers from gravitational anomalies
which render it inconsistent at the quantum level. In order to
arrive at an anomaly-free theory, we have to follow the Ho\v
rava-Witten recipe by adding boundary fields whose contribution to
the anomalies will cancel those of the gravitational theory. In
the 6D case, the available types of boundary multiplets are tensor
multiplets and hypermultiplets such as those appearing in
(\ref{e-3-11}) plus vector multiplets whose field content is given
by
\begin{equation}
\text{Vector multiplet} \quad:\quad ( A_\mu , \lambda^{-} ).
\label{e-4-1}
\end{equation}
The $D=6$, $\mathcal{N} = (0,1)$ supergravity theory coupled to
vector and tensor multiplets has been constructed in~\cite{FR} and
the inclusion of hypermultiplets has partially been obtained
in~\cite{Nishino:1997ff}. The most general up to date supergravity
coupled to vectors, tensors and hypermultiplets has been given
in~\cite{riccioni} All three types of multiplets give extra
contribution to the gravitational anomaly of the theory. Moreover,
the theory has now gauge and mixed anomalies coming from the
vector multiplets and the hypermultiplets. As we shall see in
\S\ref{sec-5}, the inclusion of these multiplets makes it possible
to cancel all anomalies of the theory via the Green-Schwarz
mechanism.

Considering the fixed plane $x_7 = 0$, we introduce $n_V$ vector
multiplets, $n_H$ hypermultiplets and $n_T$ tensor multiplets. We
take the Yang-Mills group $\mathcal{G}$ to be a product of simple
factors, $\mathcal{G} = \prod_z \mathcal{G}_z$ and we further
assume that $\mathcal{G}$ is semisimple (no $U(1)$ factors). The
vector multiplets transform in the adjoint of $\mathcal{G}$, so we
have
\begin{equation}
n_V = \dim \mathcal{G} = \sum_z \dim \mathcal{G}_z. \label{e-4-2}
\end{equation}
The hypermultiplets can also be charged under the gauge group. We
will let $n_{z,k}$ be the number of hypermultiplets in
$\mathcal{R}_{z,k}$ and $n_{zz',ij}$ be the number of
hypermultiplets in the representation
$(\mathcal{R}_{z,i},\mathcal{R}_{z',j})$ of the product group
$\mathcal{G}_z \times \mathcal{G}_{z'}$. Finally, the tensor
multiplets are not charged under $\mathcal{G}$.

Having introduced new boundary multiplets in order to achieve
anomaly cancellation, we must now determine the appropriate action
that describes these fields and their interactions with the bulk
fields. For its construction, some well-known facts about $D=6$,
$\mathcal{N}=(0,1)$ supergravity turn out to be useful. First of
all, the form of the Lagrangian is not determined by supersymmetry
alone but, instead, there are some constant factors which are
determined by anomaly cancellation conditions. Second, unlike its
counterpart in $D=10$, $\mathcal{N} = 1$ supergravity, the $D=6$
Green-Schwarz term $B_2 \wedge \tr F^2$ is not a higher-derivative
correction and hence it must be present in the low-energy action
in the first place; as we shall see, this is indeed the case.
Third, it is known that no invariant Lagrangian exists for the
case of antisymmetric tensor fields subject to self-duality
projections, and thus an action can be written down only when $n_T
= 0$; for simplicity, in this section only this case will be
considered.

In the remainder of this section, we will use the above insights
to construct, by the Noether method, the locally supersymmetric
action (up to $(\mathrm{Fermi})^4$ terms) and the supersymmetry
transformation rules (up to $(\mathrm{Fermi})^3$ terms) required
to describe the boundary multiplets and their interactions with
the bulk fields.

\subsection{Boundary vector multiplets}
\label{s-vector}

The construction of the boundary vector multiplet Lagrangian
proceeds by starting from the globally supersymmetric theory and
coupling it to gravity through the Noether method, so as to
restore local supersymmetry.

For the construction, some experience with HW theory is useful. In
that construction, it was found that the coupling of boundary
vector multiplets to the bulk supergravity leaves out certain
uncancelled supersymmetry variations, whose cancellation requires
that the field-strength $G_4$ of M-theory acquire (in the
downstairs approach) a boundary value proportional to $\tr F^2$,
where $F$ is the Yang-Mills field strength. In the present
context, we have the 3--form field strength $G_3$ obtained by
dualizing $F_4$. On these grounds, we expect that $G_3$ must
acquire a boundary value proportional to $* \tr F^2$. It turns out
that, due to some subtleties related to the duality transformation
in the presence of a boundary, this will induce the Green-Schwarz
term of the theory.

\subsubsection{The action}

We start by determining the action describing the boundary vector
multiplets and their interactions with the bulk theory, along the
lines of~\cite{Gherghetta:2002xf,Gherghetta:2002nq}. Our starting
point is the globally supersymmetric Lagrangian
\begin{equation}
e^{-1} \mathcal{L}^{(0)}_{YM} = v_z \tr_z \left( - \frac{1}{4}
e^{-\phi} F_{\mu\nu} F^{\mu\nu} - e^\phi \bar{\lambda}^i
\Gamma^\mu D_\mu \lambda_i \right). \label{e-4-4}
\end{equation}
Here, the index $z$ labels the simple factors of the gauge group,
$\tr_z$ denotes the relevant traces normalized with respect to the
fundamental representation, and $v_z$ are some numerical constants
that will be determined in the next section. The Lagrangian
(\ref{e-4-4}) is invariant under the rigid supersymmetry
transformations
\begin{equation}
\delta A_\mu = e^\phi \bar{\epsilon}^i \Gamma_\mu \lambda_i
~,\qquad \delta_0 \lambda_i = - \frac{1}{4} e^{-\phi}
\Gamma^{\mu\nu} \epsilon_i F_{\mu\nu}. \label{e-4-5}
\end{equation}
Our first step towards obtaining a locally supersymmetric theory
is to introduce the usual Noether coupling of the gravitino to the
supercurrent of the multiplet. The required term is
\begin{equation}
\mathcal{L}^{(1)}_{YM} = e v_z \tr_z \left( - \frac{1}{2}
\bar{\psi}^i_\mu \Gamma^{\nu\rho} \Gamma^\mu \lambda_i F_{\nu\rho}
\right). \label{e-4-6}
\end{equation}
Next, we must cancel the $\bar{\lambda} \Gamma F \partial \phi
\epsilon$ variation of $\mathcal{L}^{(0)}_{YM}$. This variation is
found to be
\begin{equation}
\Delta^{(1)}_{YM} = e v_z \tr_z \left( \frac{1}{4} \bar{\lambda}^i
\Gamma^{\mu\nu} \Gamma^\rho \epsilon_i F_{\mu\nu}
\partial_\rho \phi \right) \label{e-4-7}
\end{equation}
and can be cancelled by the $\delta \chi \sim \Gamma \partial \phi
\epsilon$ variation of the additional term
\begin{equation}
\mathcal{L}^{(2)}_{YM} = e v_z \tr_z \left( - \frac{1}{2}
\bar{\lambda}^i \Gamma^{\mu\nu} \chi_i F_{\mu\nu} \right).
\label{e-4-8}
\end{equation}
The introduction of these new interactions results in additional
uncancelled terms of the form $\bar{\lambda} \Gamma F G_3
\epsilon$ coming from the $\delta \psi$ and $\delta \chi$
variations of $\mathcal{L}^{(1)}_{YM}$ and
$\mathcal{L}^{(2)}_{YM}$ respectively. The first one vanishes by
the 6D identity $\Gamma^\mu \Gamma^{\nu\rho\sigma} \Gamma_\mu =
0$, while the second one is given by
\begin{equation}
\Delta^{(2)}_{YM} = e v_z \tr_z \left( \frac{1}{24 \sqrt{2}}
e^\phi \bar{\lambda}^i \Gamma^{\mu\nu} \Gamma^{\rho\sigma\tau}
\epsilon_i G_{\rho\sigma\tau} F_{\mu\nu} \right). \label{e-4-9}
\end{equation}
This can be cancelled by introducing the additional interaction
\begin{equation}
\mathcal{L}^{(3)}_{YM} = e v_z \tr_z \left( \frac{1}{12\sqrt{2}}
e^{2\phi} \bar{\lambda}^i \Gamma^{\mu\nu\rho} \lambda_i
G_{\mu\nu\rho} \right) \label{e-4-10}
\end{equation}
What remains is to cancel the $\bar{\psi} \Gamma F^2 \epsilon$ and
$\bar{\chi} \Gamma F^2 \epsilon$ terms coming from the $\delta
\lambda$ variations of $\mathcal{L}^{(1)}_{YM}$ and
$\mathcal{L}^{(2)}_{YM}$. These terms are given by
\begin{equation}
\Delta^{(3)}_{YM} = e v_z \tr_z \left( \frac{1}{8} e^{-\phi}
\bar{\psi}^i_\mu \Gamma^{\mu\nu\rho\sigma\tau} \epsilon_i
F_{\nu\rho} F_{\sigma\tau} \right), \label{e-4-11}
\end{equation}
and
\begin{equation}
\Delta^{(4)}_{YM} = e v_z \tr_z \left( \frac{1}{8} e^{-\phi}
\bar{\chi}^i \Gamma^{\mu\nu\rho\sigma} \epsilon_i F_{\mu\nu}
F_{\rho\sigma} \right). \label{e-4-12}
\end{equation}
It turns out that these terms can be cancelled by a mechanism that
will be shortly described. Provided that this happens, we can
write the locally supersymmetric Lagrangian describing the vector
multiplets and their bulk interactions in the form
\begin{eqnarray}
\label{e-4-13}
e^{-1} \mathcal{L}_{YM} = v_z \tr_z \!\!\!\!\!&\biggl(&\!\!\!\!\! - \frac{1}{4}
e^{-\phi} F_{\mu\nu} F^{\mu\nu} - e^\phi \bar{\lambda}^i
\Gamma^\mu D_\mu \lambda_i - \frac{1}{2} \bar{\psi}^i_\mu
\Gamma^{\nu\rho} \Gamma^\mu \lambda_i F_{\mu\nu} \nn\\ &&\!\!\!\!\! -
\frac{1}{2} \bar{\lambda}^i \Gamma^{\mu\nu} \chi_i F_{\mu\nu} +
\frac{1}{12\sqrt{2}} e^{2\phi} \bar{\lambda}^i \Gamma^{\mu\nu\rho}
\lambda_i G_{\mu\nu\rho} \biggr). 
\end{eqnarray}

\subsubsection{The boundary value of $G_3$
and the Green-Schwarz term}

What remains is to discuss the mechanism by which the
supersymmetry variations (\ref{e-4-11}) and (\ref{e-4-12}) can be
cancelled, which proceeds in  analogy to the HW case. There are
two ways  in doing that, the ``downstairs" and ``upstairs"
approach.
\vskip.05in \noindent {\bf Downstairs:}
 In this approach, one considers possible variations of the bulk action which can be written as total
derivatives with respect to $x_7$ (see \cite{Conrad:1997ww,Harmark:1998bs} for a
clear discussion of this point). For reasons that will soon become clear, we start by using the gamma-matrix
duality relation (\ref{a-3}) to write the uncancelled variations
(\ref{e-4-11}) and (\ref{e-4-12}) as
\begin{equation}
\Delta^{(3)}_{YM} = - \frac{1}{8}
\epsilon^{\mu\nu\rho\sigma\tau\upsilon} e^{-\phi} \bar{\psi}^i_\mu
\Gamma_\nu \epsilon_i  \tr_z \left( F_{\rho\sigma}
F_{\tau\upsilon} \right), \label{e-4-14}
\end{equation}
and
\begin{equation}
\Delta^{(4)}_{YM} = \frac{1}{16}
\epsilon^{\mu\nu\rho\sigma\tau\upsilon} e^{-\phi} \bar{\chi}^i
\Gamma_{\mu\nu} \epsilon_i  \tr_z \left( F_{\rho\sigma}
F_{\tau\upsilon} \right). \label{e-4-15}
\end{equation}
Next, we consider the $\bar{\psi} \Gamma \psi G_3$ and $\bar{\chi}
\Gamma \psi G_3$ interactions present in the \emph{bulk}
Lagrangian (\ref{e-2-34}). Starting with the former, it is easy to
see that it contains the term
\begin{equation}
- \frac{1}{24 \sqrt{2}} \int d^7 x E \sigma^2 \bar{\psi}^{Li}
\Gamma_{[L} \Gamma^{MNP} \Gamma_{Q]} \psi^Q_i G_{MNP} \to - \frac{1}{6
\sqrt{2}} \int d^7 x E \sigma^2 \bar{\psi}^{Mi} \Gamma^N \psi^P_i
G_{MNP}. \label{e-4-16}
\end{equation}
Under $\delta \psi \sim \partial \epsilon$, the variation of this
term is given by
\begin{equation}
- \frac{1}{3 \sqrt{2}} \int d^7 x E \sigma^2 \bar{\psi}^{Mi}
\Gamma^N \partial^P \epsilon_i G_{MNP}, \label{e-4-17}
\end{equation}
and its $P=7$ part contributes a total $x_7$ derivative term
which, after integration over $x_7$, results in the boundary
variation
\begin{equation}
- \frac{1}{3 \sqrt{2}} \int d^7 x \partial^7 \left( E \sigma^2
\bar{\psi}^{(7) \mu i} \Gamma^\nu \epsilon_i G_{\mu\nu 7} \right)
= - \frac{1}{3 \sqrt{2}} \int d^6 x E \sigma^2 \bar{\psi}^{(7) \mu
i} \Gamma^\nu \epsilon_i G_{\mu\nu 7}. \label{e-4-18}
\end{equation}
Repeating the same procedure for the $\bar{\chi} \Gamma \psi G_3$
interaction, we find that it contains the term
\begin{equation}
\frac{1}{6 \sqrt{10}} \int d^7 x E \sigma^2 \bar{\chi}^i \Gamma^L
\Gamma^{MNP} \psi_{Li} G_{MNP} \to \frac{1}{2 \sqrt{10}} \int d^7
x E \sigma^2 \bar{\chi}^i \Gamma^{MN} \psi^P_i G_{MNP},
\label{e-4-19}
\end{equation}
which results in the boundary variation
\begin{equation}
\frac{1}{2 \sqrt{10}} \int d^7 x \partial^7 \left( E \sigma^2
\bar{\chi}^{(7)i} \Gamma^{\mu\nu} \epsilon_i G_{\mu\nu 7} \right)
= \frac{1}{2 \sqrt{10}} \int d^6 x E \sigma^2 \bar{\chi}^{(7)i}
\Gamma^{\mu\nu} \epsilon_i G_{\mu\nu 7}. \label{e-4-20}
\end{equation}
The above are expressed in terms of the 7D fields $E$, $\sigma$,
$\psi^{(7)}_{\mu i}$ and $\chi^{(7)}_i$. To pass over to the basis
of 6D fields we use the Kaluza-Klein ansatz
(\ref{e-3-2}--\ref{e-3-6}). Then, it is easy to see that
(\ref{e-4-18}) gives rise to the 6D $\bar{\psi}^\mu \Gamma^\nu
\epsilon G_{\mu\nu 7}$ variation
\begin{equation}
\Delta^{(1)}_{B} = - \frac{1}{3 \sqrt{2}} e e^\phi \bar{\psi}^{\mu
i} \Gamma^\nu \epsilon_i G_{\mu\nu 7}, \label{e-4-21}
\end{equation}
while the combination of (\ref{e-4-18}) and (\ref{e-4-20}) results
in a 6D $\bar{\chi} \Gamma^{\mu\nu} \epsilon G_{\mu\nu 7}$
variation, given by
\begin{equation}
\Delta^{(2)}_{B} = \left( - \frac{1}{3 \sqrt{2}} \cdot
\frac{1}{10} + \frac{1}{2 \sqrt{10}} \cdot \frac{2}{\sqrt{5}}
\right) e e^\phi \bar{\chi}^i \Gamma^{\mu\nu} \epsilon_i G_{\mu\nu
7} = \frac{1}{6 \sqrt{2}} e e^\phi \bar{\chi}^i \Gamma^{\mu\nu}
\epsilon_i G_{\mu\nu 7}. \label{e-4-22}
\end{equation}
Comparing (\ref{e-4-21}) and (\ref{e-4-22}) with the uncancelled
variations (\ref{e-4-14}) and (\ref{e-4-15}) respectively, we see
that we can cancel \emph{both} of them by requiring that
$G_{\mu\nu 7}$ attain the boundary value
\begin{equation}
G^{\mu\nu 7} \bigr|_{\partial M} = - \frac{3}{4 \sqrt{2}} e^{-1}
e^{-2\phi} \epsilon^{\mu\nu\rho\sigma\tau\upsilon} v_z \tr_z
\left( F_{\rho\sigma} F_{\tau\upsilon} \right). \label{e-4-23}
\end{equation}

Moreover, we are now in a position to make sense out of our
remnant from the duality transformation of Section 2.3, namely the
surface term
\begin{equation}
S_{I,bdy} = - \frac{1}{48} \int d^7 x \epsilon^{MNPQRST}
\partial_T ( S_{MNPQ} B_{RS} ), \label{e-4-24}
\end{equation}
Its integration over $x_7$ gives rise to
the nonvanishing boundary term
\begin{equation}
S_{I,bdy} = - \frac{1}{48} \int d^6 x
\epsilon^{\mu\nu\rho\sigma\tau\upsilon} S_{\mu\nu\rho\sigma}
B_{\tau\upsilon}. \label{e-4-25}
\end{equation}
for the boundary at $x_7=0$. This implies that, for this boundary,
the algebraic equation of motion (\ref{e-2-33}) of the spacetime
components $S^{\mu\nu\rho\sigma}$ of the auxiliary field $S_4$ is
now modified to
\begin{equation}
S^{\mu\nu\rho\sigma} = \frac{1}{6} \sigma^4 E^{-1}
\epsilon^{\mu\nu\rho\sigma\tau\upsilon} \left[ G_{\tau\upsilon 7}
- 3 B_{\tau\upsilon} \delta(x_7) \right] - \frac{3}{\sqrt{2}}
\sigma^2 J^{\mu\nu\rho\sigma}. \label{e-4-26}
\end{equation}
To incorporate this modification in our theory, we must first
replace $G_{\mu\nu 7}$ by
\begin{equation}
\widetilde{G}_{\mu\nu 7} = G_{\mu\nu 7} - 3 B_{\mu\nu} \delta(x_7)
\label{e-4-27}
\end{equation}
everywhere in the 7D Lagrangian (\ref{e-2-34}) and transformation
rules (\ref{e-2-35}) and then substitute the solution
(\ref{e-4-26}) in the surface term (\ref{e-4-25}).

Let us examine the new boundary terms that arise in this way.
First, it is easy to see that the $\mathbb{Z}_2$ chirality
projections forbid the appearance of possible boundary couplings
of fermions to $B_{\mu\nu}$. Second, we notice that there also
occurs a coupling
\begin{equation}
- \frac{1}{4} \delta(0) \int d^6 x e e^{2\phi} B_{\mu\nu}
B^{\mu\nu},
 \label{e-4-28}
\end{equation}
which looks like a singular mass term for $B_{\mu\nu}$. The
appearance of such a singularity is somewhat surprising since the
theory we started with was perfectly regular; however, it is an
effect that is known to occur in dual formulations of
supergravities on orbifolds. After suitable regularization (see
\cite{Dudas:2004ni}), it turns out that this term does not affect
the mass spectrum of the theory and that $B_2$ stays massless at
tree level. Third and more important, there appears the surface
term
\begin{equation}
S_{GS} = \frac{1}{6} \int d^6 x e e^{2\phi} B_{\mu\nu} G^{\mu\nu
7}, \label{e-4-29}
\end{equation}
which, on account of (\ref{e-4-23}), results in the interaction
\begin{equation}
S_{GS} = - \frac{1}{8 \sqrt{2}} \int d^6 x
\epsilon^{\mu\nu\rho\sigma\tau\upsilon} B_{\mu\nu} v_z \tr_z
\left( F_{\rho\sigma} F_{\tau\upsilon} \right). \label{e-4-30}
\end{equation}
This is the Green-Schwarz term of our theory. Supplemented by a
suitable gravitational contribution and given an appropriate
gauge/Lorentz transformation law for $B_2$, this term can
completely cancel all anomalies of the theory. \vskip.05in
\noindent {\bf Upstairs: }To complete the discussion and to
provide a consistency check for our method, let us also briefly
describe how things work out in the ``upstairs'' approach. In this
approach, we have to go back to the 3--form theory considered
in~\cite{Gherghetta:2002xf,Gherghetta:2002nq}, where it was found
that $F_4$ should satisfy a modified Bianchi identity whose
generalization to our case reads
\begin{equation}
\partial_{[7} F_{\mu\nu\rho\sigma]} = - 3 \sqrt{2} v_z \tr_z
\left( F_{\mu\nu} F_{\rho\sigma} \right) \delta(x_7).
\label{e-4-30a}
\end{equation}
According to (\ref{e-2-26a}), this implies that the
Lagrange-multiplier action entering the duality transformation
should contain the additional term
\begin{equation}
S'_C = - \frac{1}{48} \int d^7 x
\epsilon^{\mu\nu\rho\sigma\tau\upsilon 7} B_{\mu\nu} \left[ 3
\sqrt{2} v_z \tr_z \left( F_{\rho\sigma} F_{\tau\upsilon} \right)
\delta(x_7) \right], \label{e-4-30b}
\end{equation}
besides the usual $B_2 \wedge d S_4$ term. In the ``upstairs''
approach, the $B_2 \wedge d S_4$ term can be integrated by parts
without the emergence of surface terms so that the duality
transformation works in the usual way. However, we are still left
with the extra term (\ref{e-4-30b}) which, after the trivial $x_7$
integration, exactly reproduces the Green-Schwarz interaction
(\ref{e-4-30}).

\subsection{Hypermultiplets}

To determine the action describing the boundary hypermultiplets
and their interactions with the bulk fields, we proceed as
follows. Initially, we set up the basic formalism required to
describe $D=6$ hypermultiplets and their gauging under Yang-Mills
groups. Next, we consider the simple case where the
hypermultiplets are inert under the gauge group and we construct
the appropriate locally supersymmetric action. Finally, we
consider gauging these multiplets and we determine all extra terms
required to maintain local supersymmetry.

\subsubsection{General formalism}

The construction of the gauged theory describing the
hypermultiplets was discussed in detail in \cite{Nishino:dc}. To
give a brief review, we write our $n_H$ boundary hypermultiplets
as $(\varphi^\alpha,\zeta^{a+})$ where $\alpha = 1,\ldots,4 n_H$
and $a=1,\ldots,2 n_H$. The $4 n_H$ hyperscalars $\varphi^\alpha$
parameterize a quaternionic manifold, i.e. a manifold whose
holonomy group is a subgroup of $Sp(n_H) \times Sp(1)$. We pick
this manifold to be the coset space $Sp(n_H,1) / Sp(n_H) \times
Sp(1)$ and we denote its metric by $g_{\alpha \beta} (\varphi)$. A
representative of this space can be parameterized by a matrix
$\mathbf{L}$ whose Maurer-Cartan form decomposes as
\begin{equation}
\mathbf{L}^{-1} \partial_\alpha \mathbf{L} =
\omega_\alpha^{\phantom{\alpha} ab} T_{ab}  +
\omega_\alpha^{\phantom{\alpha} ij} T_{ij} +
V_\alpha^{\phantom{\alpha} a i} T_{a i}, \label{e-4-31}
\end{equation}
where $T_{ab}$ and $T_{ij}$ are the $Sp(n_H)$ and $Sp(1)$
generators, $\omega_\alpha^{\phantom{\alpha} ab}$ and
$\omega_\alpha^{\phantom{\alpha} ij}$ are the associated
connections and $T_{a i}$ and $V_\alpha^{\phantom{\alpha} a i}$
are the coset generators and vielbeins. The $Sp(n_H)$ and $Sp(1)$
curvatures are denoted by
$\Omega_{\alpha\beta}^{\phantom{\alpha\beta} a b}$ and
$\Omega_{\alpha\beta}^{\phantom{\alpha\beta} i j}$ respectively
and the latter is expressed in terms of the vielbeins as
\begin{equation}
\Omega_{\alpha\beta}^{\phantom{\alpha\beta} i j} = 2 \left(
V_{\alpha a}^{\phantom{\alpha a} i} V_\beta^{\phantom{\beta} a j}
+ V_{\alpha a}^{\phantom{\alpha a} j} V_\beta^{\phantom{\beta} a
i} \right). \label{e-4-32}
\end{equation}

To gauge the hypermultiplets, we must make a choice for the
Yang-Mills group $\mathcal{G} = \prod_z \mathcal{G}_z$ and the
representations in which the hypermultiplets transform. One can
also take some of the $\mathcal{G}_z$'s to be subgroups of the
$Sp(n_H) \times Sp(1)$ holonomy group whose $Sp(1)$ factor is
identified with the R-symmetry group. In this section, we will
take $\mathcal{G}_1 = Sp(n_H)$ and $\mathcal{G}_2 = Sp(1)$,
without further specification of the remaining factors. The
corresponding gauge fields are denoted by $A_\mu^{ab}$ and
$A_\mu^{ij}$ and their field strengths are defined in the usual
way. Under a $Sp(n_H) \times Sp(1)$ gauge transformation, the
gauge fields and the hyperscalars transform according to
\begin{equation}
\delta A_\mu^{ab} = D_\mu \Lambda^{ab} ~,\qquad \delta
A_\mu^{ij} = D_\mu \Lambda^{ij}, \label{e-4-33}
\end{equation}
and
\begin{equation}
\delta \varphi^\alpha = \Lambda^{ab} \xi^\alpha_{\phantom{\alpha}
ab} + \Lambda^{ij} \xi^\alpha_{\phantom{\alpha} ij},
\label{e-4-34}
\end{equation}
where $\xi^\alpha_{\phantom{\alpha} ab}$ and
$\xi^\alpha_{\phantom{\alpha} ij}$ are $Sp(n_H)$ and $Sp(1)$
Killing vectors, given by
\begin{equation}
\xi^\alpha_{\phantom{\alpha} ab} = T_{ab} \varphi^\alpha
~,\qquad \xi^\alpha_{\phantom{\alpha} ij} = T_{ij}
\varphi^\alpha. \label{e-4-35}
\end{equation}
The covariant derivative acting on the hyperscalars is then
\begin{equation}
\mathcal{D}_\mu \varphi^\alpha = \partial_\mu \varphi^\alpha -
A_\mu^{ab} \xi^\alpha_{\phantom{\alpha} ab} - A_\mu^{ij}
\xi^\alpha_{\phantom{\alpha} ij}. \label{e-4-36}
\end{equation}
As for the spinor covariant derivatives, they are defined by
adding the appropriate composite connections of the form
$\mathcal{D}_\mu \varphi^\alpha \omega_\alpha$ plus the gauge
field terms. In particular, the covariant derivative of the spinor
$\epsilon^i$ parameterizing the SUSY transformations is modified
to
\begin{equation}
\mathcal{D}_\mu \epsilon^i = D_\mu \epsilon^i + \left(
\mathcal{D}_\mu \varphi^\alpha \right)
\omega_\alpha^{\phantom{\alpha} ij} \epsilon_j + A_\mu^{ij}
\epsilon_j. \label{e-4-37}
\end{equation}
As a result, the commutator of two covariant derivatives on
$\epsilon^i$ is given by
\begin{equation}
[ \mathcal{D}_\mu , \mathcal{D}_\nu ] \epsilon^i = \frac{1}{4}
R_{\mu\nu\rho\sigma} \Gamma^{\rho\sigma} \epsilon^i +
\mathcal{D}_\mu \varphi^\alpha \mathcal{D}_\nu \varphi^\beta
\Omega_{\alpha\beta}^{\phantom{\alpha\beta} ij} \epsilon_j - \tr_z
(F_{\mu\nu} \C^{ij} \epsilon_j), \label{e-4-38}
\end{equation}
where $\C^{ij}$ denotes the following triplet of $Sp(n_H) \times
Sp(1)$ matrices
\begin{equation}
\C^{ij} = \omega_\alpha^{\phantom{\alpha} ij}
\xi^\alpha_{\phantom{\alpha} cd} T^{cd} +
\omega_\alpha^{\phantom{\alpha} ij} \xi^\alpha_{\phantom{\alpha}
kl} T^{kl} - T^{ij}. \label{e-4-39}
\end{equation}
These matrices satisfy the identity \cite{Nishino:dc}
\begin{equation}
\mathcal{D}_\mu \C^{ij} = \left( \mathcal{D}_\mu \varphi^\alpha
\right) \Omega_{\alpha\beta}^{\phantom{\alpha\beta} ij} \xi^\beta,
\label{e-4-40}
\end{equation}
where
\begin{equation}
\xi^\alpha = \xi^{\alpha ab} T_{ab} + \xi^{\alpha ij} T_{ij}.
\label{e-4-41}
\end{equation}

The above construction is purely six-dimensional. To implement it in the context of our bulk-brane theory, we must find a 7D explanation for the quaternionic structure of the scalar manifold, or else our approach will not be consistent with local supersymmetry in 6D. We recall that, in the 6D case, the quaternionic structure is a result of the fact that the gravitino is charged under the $Sp(1)$ 6D R-symmetry so that the supersymmetry variation of its kinetic term gives rise to a term involving the $Sp(1)$ curvature. However, in our 7D theory, the gravitino is a bulk field which does not couple to the boundary $Sp(1)$ connection and the quaternionic structure is not a priori imposed. However, we will soon see that the required structure does indeed arise by a boundary condition on the bulk gauge field.

\subsubsection{The action for neutral hypermultiplets}

To construct the action, let us begin from the ungauged theory.
Our starting point is the globally supersymmetric Lagrangian
\begin{equation}
e^{-1} \mathcal{L}^{(0)}_{H} = - g_{\alpha\beta}(\varphi)
\partial_\mu \varphi^\alpha \partial^\mu \varphi^\beta -
\bar{\zeta}^a \Gamma^\mu D_\mu \zeta_a, 
\label{e-4-42}
\end{equation}
which is invariant under the transformations
\begin{equation}
\label{e-4-43}
\delta \varphi^\alpha = V^\alpha_{\phantom{\alpha} a i}
\bar{\zeta}^a \epsilon^i ~,\qquad \delta \zeta^a =
V_\alpha^{\phantom{\alpha} a i} \Gamma^\mu \partial_\mu
\varphi^\alpha \epsilon^i. 
\end{equation}
As before, we introduce again the appropriate interaction of the
gravitino with the supercurrent,
\begin{equation}
\mathcal{L}^{(1)}_{H} = 2 e \bar{\psi}^i_\mu \Gamma^\nu \Gamma^\mu
\zeta^a V_{\alpha a i} \partial_\nu \varphi^\alpha, \label{e-4-44}
\end{equation}
which, under $\delta \psi \sim \Gamma G \epsilon$, yields the
uncancelled term
\begin{equation}
\Delta^{(1)}_{H} = - \frac{1}{6 \sqrt{2}} e e^\phi
\bar{\epsilon}^i \Gamma^\mu \Gamma^{\nu\rho\sigma} \zeta^a
V_{\alpha a i} \partial_\mu \varphi^\alpha G_{\nu\rho\sigma}.
\label{e-4-45}
\end{equation}
This term cancels if we introduce the interaction
\begin{equation}
\mathcal{L}^{(2)}_{H} = - \frac{1}{12 \sqrt{2}} e e^\phi
\bar{\zeta}^a \Gamma^{\mu\nu\rho} \zeta_a G_{\mu\nu\rho}.
\label{e-4-46}
\end{equation}
Finally, there is also a term arising from the $\delta \zeta$
variation in $\mathcal{L}^{(1)}_{H}$, which is given by
\begin{equation}
\Delta^{(2)}_{H} = - \frac{1}{2} e \bar{\psi}_{\mu i}
\Gamma^{\mu\nu\rho} \partial_\nu \varphi^\alpha \partial_\rho
\varphi^\beta \Omega_{\alpha\beta}^{\phantom{\alpha\beta} ij}
\epsilon_j. \label{e-4-47}
\end{equation}
To cancel this term, we note, as in~\cite{Gherghetta:2002xf,Gherghetta:2002nq}, that the bulk theory contains a  $\bar{\psi} \Gamma \psi F_2$ interaction. Under $\delta \psi \sim \partial \epsilon$, this term has the variation
\begin{equation}
- \frac{i}{2 \sqrt{2}} E \sigma \bar{\psi}_{L i}
\Gamma^{LMNP} \partial_P \epsilon_j F_{MN}^{\phantom{MN} ij}, \label{e-4-a1}
\end{equation}
where $F_{MN \phantom{i} j}^{\phantom{MN} i} = F^I_{MN} L_{I \phantom{i} j}^{\phantom{I} i}$. Following the same reasoning as in Section \ref{s-vector}, we take the total-derivative contribution of the $P=7$ part, integrate over $x_7$ and express the result in the 6D basis. The result is
\begin{equation}
\Delta^{(3)}_{H} = - \frac{i}{2 \sqrt{2}} e e^{-\phi/2} \bar{\psi}_{\mu i}
\Gamma^{\mu\nu\rho} F_{\nu\rho}^{\phantom{\nu\rho} ij} \epsilon_j. \label{e-4-a2}
\end{equation}
This can exactly cancel $\Delta^{(2)}_{H}$ provided that $F_{\mu\nu}^{\phantom{\mu\nu} ij}$ (normally vanishing on the boundary) is subject to the following boundary condition, in the downstairs approach
\begin{equation}
F_{\mu\nu}^{\phantom{\mu\nu} ij} \bigr|_{\partial M} = i \sqrt{2} e^{\phi/2} \partial_\mu \varphi^\alpha \partial_\nu \varphi^\beta \Omega_{\alpha\beta}^{\phantom{\alpha\beta} ij},
\label{e-4-a3}
\end{equation}
or the corresponding Bianchi identity in the upstairs approach. As explained in detail in~\cite{Gherghetta:2002nq}, this does indeed induce the required quaternionic structure on the scalar manifold as a result of the Bianchi identity $D_{[\lambda} F_{\mu\nu]}^{\phantom{\mu\nu} ij} = 0$.

Collecting all terms, we arrive at the locally supersymmetric action for neutral hypermultiplets,
\begin{eqnarray}
\label{e-4-48}
e^{-1} \mathcal{L}_{H} &=& - g_{\alpha\beta}(\varphi)
\partial_\mu \varphi^\alpha \partial^\mu \varphi^\beta -
\bar{\zeta}^a \Gamma^\mu D_\mu \zeta_a + 2 \bar{\psi}^i_\mu
\Gamma^\nu \Gamma^\mu \partial_\nu \varphi^\alpha V_{\alpha a i}
\zeta^a \nn \\ &&
- \frac{1}{12 \sqrt{2}} e e^\phi \bar{\zeta}^a
\Gamma^{\mu\nu\rho} \zeta_a G_{\mu\nu\rho}. 
\end{eqnarray}

\subsubsection{Gauging}

Next, we shall extend our results for the case where the
hypermultiplets are charged under the gauge group. The first step
in the construction of the theory is to replace all derivatives by
covariant ones with respect to the gauge group, that is, replace
the hypermultiplet Lagrangian and transformation rules by
\begin{eqnarray}
\label{e-4-49}
e^{-1} \mathcal{L}_{H} &=& - g_{\alpha\beta}(\varphi)
\mathcal{D}_\mu \varphi^\alpha \mathcal{D}^\mu \varphi^\beta -
\bar{\zeta}^a \Gamma^\mu \mathcal{D}_\mu \zeta_a + 2
\bar{\psi}^i_\mu \Gamma^\nu \Gamma^\mu \mathcal{D}_\nu
\varphi^\alpha V_{\alpha a i} \zeta^a \nn \\ && - \frac{1}{12
\sqrt{2}} e e^\phi \bar{\zeta}^a \Gamma^{\mu\nu\rho} \zeta_a
G_{\mu\nu\rho} , 
\end{eqnarray}
and
\begin{equation}
\delta \varphi^\alpha = V^\alpha_{\phantom{\alpha} a i}
\bar{\zeta}^a \epsilon^i~,\qquad \delta \zeta^a =
V_\alpha^{\phantom{\alpha} a i} \Gamma^\mu \mathcal{D}_\mu
\varphi^\alpha \epsilon^i . \label{e-4-50}
\end{equation}
After this modification, there arise additional uncancelled terms. First of all, the variation $\Delta^{(2)}_{H}$ is replaced by its covariant version. To cancel it, we could think of modifying the boundary condition (\ref{e-4-a3}) to involve covariant derivatives as well. However, upon applying this naive modification, we find that  $\mathcal{D}_{[\lambda} F_{\mu\nu]}^{\phantom{\mu\nu} ij}$ contains a term proportional to $\tr_z ( \Omega_{\alpha\beta}^{\phantom{\alpha\beta} ij} \mathcal{D}_\mu \varphi^\alpha F_{\lambda\nu} \xi^\beta )$ and thus $F_{\mu\nu}^{\phantom{\mu\nu} ij}$ fails to satisfy the 6D Bianchi identity required for the quaternionic structure. The correct modification to $\ref{e-4-a3}$ is instead given by
\begin{equation}
F_{\mu\nu}^{\phantom{\mu\nu} ij} \bigr|_{\partial M} = i \sqrt{2} e^{\phi/2} \left[ \mathcal{D}_\mu \varphi^\alpha \mathcal{D}_\nu \varphi^\beta \Omega_{\alpha\beta}^{\phantom{\alpha\beta} ij} - \tr_z
(F_{\mu\nu} \C^{ij} ) \right].
\label{e-4-a4}
\end{equation}
and, by virtue of (\ref{e-4-40}), the extra term on the RHS restores the Bianchi identity maintaining the quaternionic structure; as we shall see below, this extra term is actually necessary for local supersymmetry. After imposing (\ref{e-4-a4}), (\ref{e-4-a2}) is replaced by the additional variations
\begin{equation}
\Delta^{(1)}_G = \frac{1}{2} e \bar{\psi}_{\mu i}
\Gamma^{\mu\nu\rho} \mathcal{D}_\nu \varphi^\alpha
\mathcal{D}_\rho \varphi^\beta
\Omega_{\alpha\beta}^{\phantom{\alpha\beta} ij} \epsilon_j,
\label{e-4-51}
\end{equation}
and
\begin{equation}
\Delta^{(2)}_G = \frac{1}{2} e  \tr_z \left( \C_{ij}
\bar{\psi}^i_{\mu} \Gamma^{\mu\nu\rho} \epsilon^j F_{\nu\rho}
\right), \label{e-4-52}
\end{equation}
the first of which exactly cancels the covariant version of $\Delta^{(2)}_{H}$. Meanwhile, the hyperino kinetic term also gives rise to the extra variation
\begin{equation}
\Delta^{(3)}_G = - e  \tr_z \left( \bar{\zeta}_a \Gamma^{\mu\nu}
F_{\mu\nu} \epsilon_i V_\alpha^{\phantom{\alpha} a i} \xi^\alpha
\right) . \label{e-4-53}
\end{equation}
To cancel $\Delta^{(2)}_G$, we may either modify the gaugino SUSY
transformation law by adding the extra term
\begin{equation}
\delta_1 \lambda_i = a v_z^{-1} \C_{ij} \epsilon^j, \label{e-4-54}
\end{equation}
(summation over $z$ implicit) or introduce the additional
interaction
\begin{equation}
\mathcal{L}^{(1)}_G = b e e^\phi \tr_z \left( \C_{ij}
\bar{\psi}_\mu^i \Gamma^\mu \lambda^j \right). \label{e-4-55}
\end{equation}
Here, $a$ and $b$ are two coefficients, which can be determined by
considering the $\bar{\psi} \Gamma F \epsilon$ terms. The
variations of this type arising from the $\delta_1 \lambda$
variation of $\mathcal{L}_{YM}$ is
\begin{equation}
\Delta^{(4)}_G = \frac{a}{2} e  \tr_z \left( \C_{ij}
\bar{\psi}^i_{\mu} \Gamma^{\mu\nu\rho} \epsilon^j F_{\nu\rho}
\right) - a e  \tr_z \left( \C_{ij} \bar{\psi}^{\mu i} \Gamma^\nu
\epsilon^j F_{\mu\nu} \right), \label{e-4-56}
\end{equation}
while the $\delta_0 \lambda$ variation of $\mathcal{L}^{(1)}_G$
gives
\begin{equation}
\Delta^{(5)}_G = - \frac{b}{4} e  \tr_z \left( \C_{ij}
\bar{\psi}^i_{\mu} \Gamma^{\mu\nu\rho} \epsilon^j F_{\nu\rho}
\right) - \frac{b}{2} e  \tr_z \left( \C_{ij} \bar{\psi}^{\mu i}
\Gamma^\nu \epsilon^j F_{\mu\nu} \right). \label{e-4-57}
\end{equation}
We observe that the requirement for cancellation of the
$\bar{\psi}_\mu \Gamma^{\mu\nu\rho} F_{\nu\rho} \epsilon$ and
$\bar{\psi}^\mu \Gamma^\nu F_{\mu\nu} \epsilon$ terms fixes the
coefficients $a$ and $b$ to
\begin{equation}
a = -\frac{1}{2} ~,\qquad b = 1 . \label{e-4-58}
\end{equation}
Next, let us consider the $\delta_1 \lambda$ variation of the
gaugino kinetic term in $\mathcal{L}_{YM}$. Performing an
integration by parts and using (\ref{e-4-40}), we obtain
\begin{eqnarray}
\label{e-4-59}
\Delta^{(6)}_G &=& e e^\phi \tr_z \left( \C_{ij} \bar{\lambda}^i
\Gamma^\mu \mathcal{D}_\mu \epsilon^j \right) + e e^\phi  \tr_z
\left( \bar{\lambda}_i \Gamma^\mu
\Omega_{\alpha\beta}^{\phantom{\alpha\beta} ij}
\mathcal{D}_\mu \varphi^\alpha \xi^\beta \epsilon_j \right) \nn\\
&&- \frac{1}{2} e e^\phi \tr_z \left( \C_{ij} \bar{\lambda}^i
\Gamma^\mu \epsilon^j
\partial_\mu \phi \right). 
\end{eqnarray}
On the other hand, using $\Gamma^\mu \Gamma^{\nu\rho\sigma}
\Gamma_\mu = 0$, we find that the $\delta \psi$ variation of
$\mathcal{L}^{(1)}_G$ is given by
\begin{equation}
\Delta^{(7)}_G = - e e^\phi \tr_z \left( \C_{ij} \bar{\lambda}^i
\Gamma^\mu \mathcal{D}_\mu \epsilon^j  \right), \label{e-4-60}
\end{equation}
and it exactly cancels the first term of $\Delta^{(6)}_G$. The
second term can be cancelled by the $\delta \zeta$ variation of
the new term
\begin{equation}
\mathcal{L}^{(2)}_G = - 4 e e^\phi \tr_z \left( \bar{\lambda}_i
\zeta_a V_\alpha^{\phantom{\alpha} a i} \xi^\alpha \right) .
\label{e-4-61}
\end{equation}
Taking the $\delta_0 \lambda$ variation of this term, we see that $\Delta^{(3)}_G$ cancels as well. We thus confirm the fact that the boundary condition (\ref{e-4-a4}) that induces the variation $\Delta^{(6)}_G$ and necessitates the addition of $\mathcal{L}^{(2)}_G$ is necessary for local supersymmetry; this serves to emphasize that local 6D supersymmetry requires the scalar manifold to be quaternionic. Two other uncancelled terms are the $\delta_1 \lambda$ variations of the $\bar{\lambda} \Gamma \chi F_2$ and $\bar{\lambda} \Gamma \lambda G_3$ terms of $\mathcal{L}_{YM}$, given by
\begin{equation}
\Delta^{(8)}_G = \frac{1}{4} e  \tr_z \left( \C_{ij}
\bar{\chi}^i \Gamma^{\mu\nu} \epsilon^j F_{\mu\nu} \right),
\label{e-4-62}
\end{equation}
and
\begin{equation}
\Delta^{(9)}_G = - \frac{1}{12\sqrt{2}} e e^{2\phi} \tr_z \left(
\C_{ij} \bar{\lambda}^i \Gamma^{\mu\nu\rho} \epsilon^j
G_{\mu\nu\rho} \right), \label{e-4-63}
\end{equation}
respectively. $\Delta^{(8)}_G$ is cancelled by the $\delta_0
\lambda$ variation of yet another new term
\begin{equation}
\mathcal{L}^{(3)}_G = e e^\phi \tr_z \left( \C_{ij} \bar{\chi}^i
\lambda^j \right), \label{e-4-64}
\end{equation}
whose $\delta \chi$ variation is given by
\begin{equation}
\Delta^{(9)}_G = - \frac{1}{2} e e^\phi \tr_z \left( \C_{ij}
\bar{\lambda}^i \Gamma^\mu \epsilon^j \partial_\mu \phi \right) +
\frac{1}{12 \sqrt{2}} e e^{2\phi} \tr_z \left( \C_{ij}
\bar{\lambda}^i \Gamma^{\mu\nu\rho} \epsilon^j G_{\mu\nu\rho}
\right), \label{e-4-65}
\end{equation}
so that its first part cancels the third term of $\Delta^{(6)}_G$
and its second part cancels $\Delta^{(9)}_G$. What remains to be
cancelled are the $\delta_1 \lambda$ variations of
$\mathcal{L}^{(1)}_G$ and $\mathcal{L}^{(3)}_G$, given by
\begin{equation}
\Delta^{(10)}_G = - \frac{1}{2} e e^\phi v_z^{-1} \tr_z \left(
\C_{ij} \C^{jk} \bar{\psi}^i_{\mu} \Gamma^\mu \epsilon_k \right) ,
\label{e-4-66}
\end{equation}
and
\begin{equation}
\Delta^{(11)}_G = - \frac{1}{2} e e^\phi v_z^{-1} \tr_z \left(
\C_{ij} \C^{jk} \bar{\chi}^i \epsilon_k \right), \label{e-4-67}
\end{equation}
respectively. To cancel them, we introduce the term
\begin{equation}
\mathcal{L}^{(4)}_G = - \frac{1}{2} e e^\phi  v_z^{-1} \tr_z
\left( \C_{ij} \C^{ij} \right), \label{e-4-68}
\end{equation}
whose SUSY variation is given by
\begin{equation}
\Delta^{(12)}_G = - \frac{1}{4} e e^\phi v_z^{-1} \tr_z \left(
\C_{ij} \C^{ij} \bar{\psi}^k_{\mu} \Gamma^\mu \epsilon_k \right) -
\frac{1}{4} e e^\phi v_z^{-1} \tr_z \left( \C_{ij} \C^{ij}
\bar{\chi}^k \epsilon_k \right), \label{e-4-69}
\end{equation}
and the desired cancellation does indeed occur, due to the spinor
identity
\begin{equation}
\bar{\psi}^i \chi_j = - \frac{1}{2} \delta^i_j \bar{\psi}^k \chi_k
+ \frac{1}{2} (\sigma_I)^i_{\phantom{i} j} [\bar{\psi} \chi]_I
\qquad;\qquad [\bar{\psi} \chi]_I \equiv (\sigma_I)^i_{\phantom{i}
j} \bar{\psi}^j \chi_i. \label{e-4-70}
\end{equation}
To summarize, the terms that should be added to
$\mathcal{L}_{bulk} + \mathcal{L}_{0} + \mathcal{L}_{YM} +
\mathcal{L}_{GS} + \mathcal{L}_{H}$ in order to restore local
supersymmetry are the following
\begin{equation}
e^{-1} \mathcal{L}_G = e^\phi \tr_z \left( \C_{ij} \bar{\psi}_\mu^i
\Gamma^\mu \lambda^j + \C_{ij} \bar{\chi}^i \lambda^j - 4
\bar{\lambda}_i \zeta_a V_\alpha^{\phantom{\alpha} a i} \xi^\alpha
- \frac{1}{2} v_z^{-1} \C_{ij} \C^{ij} \right). \label{e-4-71}
\end{equation}

\section{Anomaly cancellation}
\label{sec-5}

In this section, we shall describe the mechanism by which the
anomalies of the bulk theory cancel after introducing additional
multiplets living on the boundary. We begin by analyzing the
gravitational anomalies of the bulk theory plus the gravitational,
gauge and mixed anomalies arising due to the extra boundary
multiplets. Next, we present the conditions necessary for the
cancellation of anomalies, which result in stringent constraints on
the boundary matter content. Finally, we present the Green-Schwarz
mechanism employed for local anomaly cancellation and we briefly
comment on the issues of global anomaly cancellation and
non-perturbative anomalies.

\subsection{Anomaly analysis}

Here we will analyze the anomaly structure of the bulk-boundary theory and we will derive the anomaly cancellation conditions that must be satisfied by any consistent model. For the sake of simplicity, we restrict to the case where the gauge group is contained in $Sp(n_H)$, i.e. does not include a subgroup of the $Sp(1)$ R-symmetry group of the 6D SUSY algebra.

As remarked earlier on, the bulk theory dimensionally reduced on
${\bf S}^1/\mathbb{Z}_2$ has gravitational anomalies as an obvious
consequence of its chiral spectrum. By standard arguments, the
anomalies should be equally distributed in the two fixed planes.
To determine their form, we first observe that the contributions
coming from the self-dual and anti-self-dual parts of $B_2$ cancel
each other. Therefore, the anomaly on each fixed plane $i$ is half
of that corresponding to a negative-chirality gravitino and $N+2$
positive-chirality spinors. Including another factor of
$\frac{1}{2}$ due to the symplectic Majorana-Weyl property of the
fermions, we find
\begin{equation}
I^{bulk}_{8} (R) = \frac{1}{4} \left[ - I^{3/2}_{8} (R) + (N+2)
I^{1/2}_{8} (R) \right]. \label{e-5-1}
\end{equation}
or, using the explicit expressions for the anomaly polynomials
\cite{Alvarez-Gaume:1983ig} summarized in Appendix \ref{appb},
\begin{equation}
I^{bulk}_{8} (R) = \frac{1}{960} \left[ (243 - N) \tr R^4 -
\frac{5}{4} (45 + N) (\tr R^2)^2 \right]. \label{e-5-2}
\end{equation}

Next, let us consider the gravitational anomaly of the boundary
multiplets on a given fixed plane. The inclusion of the boundary
multiplets, namely the $n_T$ tensor multiplets, the $n_H$
hypermultiplets and the $n_V$ vector multiplets, leads to the
following contribution to the gravitational anomaly
\begin{equation}
I^{bdy}_{8} (R) = \frac{1}{2} \left[ n_T I^{A}_{8} (R) + \left(
n_T + n_H - n_V \right) I^{1/2}_{8} (R) \right], \label{e-5-3}
\end{equation}
which has the explicit form
\begin{equation}
I^{bdy}_{8} (R) = \frac{1}{960} \left[ \left( 2 n_{V} - 2 n_{H} -
58 n_{T} \right) \tr R^4 + \frac{5}{4} \left( 2 n_{V} - 2 n_{H} +
14 n_{T} \right) (\tr R^2)^2 \right]. \label{e-5-4}
\end{equation}
Putting everything together, we find that the total gravitational
anomaly is given by
\begin{eqnarray}
\label{e-5-5}
I^{total}_{8} (R) = \frac{1}{960} &\biggl[& \left( 2 n_{V} - 2
n_{H} - 58 n_{T} + 243 - N \right) \tr R^4 \nn\\ && + \frac{5}{4}
\left( 2 n_{V} - 2 n_{H} + 14 n_{T} - 45 - N \right) (\tr R^2)^2
\biggr], 
\end{eqnarray}
Since $SO(5,1)$ has an independent fourth-order Casimir, the $\tr
R^4$ term in the above anomaly is irreducible and its coefficient
is required to vanish. Hence, we must require that
\begin{equation}
2 n_{H} + 58 n_{T} - 2 n_{V} = 243 - N, \label{e-5-6}
\end{equation}
in which case the total gravitational anomaly is given by the
expression
\begin{equation}
I_{8} (R) = - \frac{3}{8} \left( 1 - \frac{n_{T}}{4} \right) (\tr
R^2)^2 \equiv - \frac{3}{8} k (\tr R^2)^2. \label{e-5-7}
\end{equation}

After the inclusion of the boundary multiplets, the theory also
has gauge and mixed anomalies arising from the couplings of
spinors to the gauge field. The relevant contributions come from
the negative-chirality gauginos of the vector multiplet and from
the positive-chirality hyperinos. To study these anomalies, we will
denote the trace in a generic representation $\mathcal{R}_{z,k}$
of $\mathcal{G}_z$ by $\tr_{z,k}$ while, as usual, we will reserve
the notation $\Tr_z$ for the trace in the adjoint. Also, following
Schwarz \cite{Schwarz:1995zw}, we will define
\begin{equation}
X^{(n)}_z = \Tr_z F^n - \sum_k n_{z,k} \tr_{z,k} F^n ~,\qquad
Y_{zz'} = \sum_{i,j} n_{zz',ij} \tr_{z,i} F^2 \tr_{z',j} F^2.
\label{e-5-8}
\end{equation}
Starting from gauge anomalies, we find that the anomaly
corresponding to fermions transforming in representations of a
single $\mathcal{G}_z$ factor is given by
\begin{eqnarray}
\label{e-5-9}
I_{8,z} (F) &=& \frac{1}{2} \left[ - I_{8,z}^{1/2} (F) + \sum_k
n_{z,k} I_{8,z,k}^{1/2} (F) \right]
\nn \\  &=& \frac{1}{2} \left( \Tr_z F^4
- \sum_k n_{z,k} \tr_{z,k} F^4 \right) = \frac{1}{2} X^{(4)}_z,
\end{eqnarray}
while the anomaly corresponding to fermions transforming in
representations of $\mathcal{G}_z \times \mathcal{G}_{z'}$ is
\begin{equation}
I_{8,zz'} (F) = - 3 \sum_{i,j} n_{zz',ij} \tr_{z,i} F^2 \tr_{z',j}
F^2 = - 3 Y_{zz'}. \label{e-5-10}
\end{equation}
Summing over the various gauge group factors, we find the gauge
anomaly
\begin{equation}
I_{8} (F) = \frac{1}{2} \sum_z X^{(4)}_z - \frac{3}{2} \sum_{z \ne
z'} Y_{zz'}. \label{e-5-11}
\end{equation}
where the extra $\frac{1}{2}$ in front of $Y_{zz'}$ takes care of
double counting. Similarly, for the mixed anomaly, the
contribution from $\mathcal{G}_z$ is given by
\begin{equation}
I_{8,z} (F,R) = - \frac{1}{8} \tr R^2 \left( \Tr_z F^2 - \sum_k
n_{z,k} \tr_{z,k} F^2 \right) = - \frac{1}{8} \tr R^2 X^{(2)}_z.
\label{e-5-12}
\end{equation}
and the sum over all gauge group factors reads
\begin{equation}
I_{8} (F,R) = - \frac{1}{8} \tr R^2 \sum_z X^{(2)}_z.
\label{e-5-13}
\end{equation}
Collecting all contributions, we finally arrive at the total
anomaly
\begin{equation}
I_{8} = \frac{1}{2} \left[ \sum_z X^{(4)}_z - 3 \sum_{z \ne z'}
Y_{zz'} - \frac{1}{4} \tr R^2 X^{(2)}_z - \frac{3 k}{4} (\tr
R^2)^2 \right]. \label{e-5-14}
\end{equation}

In order for this anomaly to cancel via the Green-Schwarz
mechanism, the polynomial (\ref{e-5-14}) must factorize. To
determine the conditions under which factorization can occur, it
is convenient to express the $F$--dependent terms in terms of
traces with respect to the fundamental representation of each
$\mathcal{G}_z$. We write
\begin{eqnarray}
\label{e-5-15}
X^{(4)}_z &=& \alpha_z \tr_z F^4 + \gamma_z ( \tr_z F^2 )^2 \nn\\
X^{(2)}_z &=& \beta_z \tr_z F^2 \\
Y_{zz'} &=& \delta_{zz'} \tr_z F^2 \tr_{z'} F^2 \nn 
\end{eqnarray}
where $\alpha_z$, $\beta_z$, $\gamma_z$ and $\delta_{zz'}$ are
some coefficients which depend on the various groups and
representations. Substituting into (\ref{e-5-14}), we write our
anomaly polynomial as
\begin{eqnarray}
\label{e-5-16}
I_{8} = \frac{1}{2} &\biggl[& \sum_z \alpha_z \tr_z F^4 + \sum_z
\gamma_z \left( \tr_z F^2 \right)^2 - 3 \sum_{z \ne z'}
\delta_{zz'} \tr_z F^2 \tr_{z'} F^2 \nn\\ && - \frac{1}{4} \tr R^2
\sum_z \beta_z \tr_z F^2 - \frac{3}{4} (\tr R^2)^2 \biggr].
\end{eqnarray}
Since each $\tr_z F^4$ term carries a coefficient $\alpha_z$
(generally dependent on the $n_{z,k}$'s), factorization can occur
either (i) when every $\tr_z F^4$ term is reducible, so that
$\alpha_z$ is manifestly zero or (ii) when we \emph{set} $\alpha_z
= 0$ by our choice of boundary multiplets. The first condition is
strictly group-theoretical and holds for all representations
having no fourth-order Casimirs, while the second one is
model-dependent. In any case, our anomaly polynomial can be
written as
\begin{eqnarray}
\label{e-5-17}
I_{8} = - \frac{3}{8} &\biggl[& k (\tr R^2)^2 + \frac{1}{3} \tr
R^2 \sum_z \beta_z \tr_z F^2 - \frac{4}{3} \sum_z \gamma_z \left(
\tr_z F^2 \right)^2 \nn \\ && + 4 \sum_{z \ne z'} \delta_{zz'}
\tr_z F^2 \tr_{z'} F^2 \biggr]. 
\end{eqnarray}

In order for this to cancel via the Green-Schwarz mechanism, it
must factorize as
\begin{equation}
I_{8} = - \frac{3k}{8} \left( c_z \tr_z F^2 - \tr R^2 \right)
\left( \tilde{c}_z \tr_z F^2 - \tr R^2 \right).
 \label{e-5-18}
\end{equation}
where we reintroduced the summation convention for $z$. This can
happen provided that (i) for each $z$,
\begin{equation}
c_z + \tilde{c}_z = - \frac{1}{3k} \beta_z ~,\qquad c_z
\tilde{c}_z = - \frac{4}{3k} \gamma_z \label{e-5-19}
\end{equation}
and (ii) for each pair $z \ne z'$,
\begin{equation}
c_z \tilde{c}_{z'} + c_{z'} \tilde{c}_z = \frac{4}{k} \delta_{zz'}
\label{e-5-20}
\end{equation}
Eqs. (\ref{e-5-19}) and (\ref{e-5-20}) are similar to the
conditions found in \cite{Erler:1993zy} and they are interpreted
as follows. The first two conditions of (\ref{e-5-19}) can be used
to determine $c_z$ and $\tilde{c}_z$: real solutions exist when
\begin{equation}
\beta_z^2 + 48 k \gamma_z \ge 0, \label{e-5-21}
\end{equation}
in which case they are given by the roots of the equation
\begin{equation}
3 k x^2 + \beta_z x - 4 \gamma_z = 0. \label{e-5-22}
\end{equation}
The third condition (\ref{e-5-20}) amounts then to a set of
non-trivial relations that must be satisfied by the
group-theoretical coefficients $\beta_z$, $\gamma_z$ and
$\delta_{zz'}$, which, in turn result to stringent restrictions on
the boundary matter content.

The extension of the above in the case where a $Sp(1)$ or $U(1)$ R-symmetry subgroup is gauged is straightforward. One has just to supplement the gauge and mixed anomaly polynomials by the contributions from R-charged fermions, i.e. the boundary gauginos (and possibly tensorinos). One again arrives at similar expressions as those presented just above.

So far, we have considered only one single fixed plane. However,
having in mind that we intend to cancel the anomalies by a
Green-Schwarz mechanism using a \emph{single} bulk $2$--form, we
have to ensure that one of the two factors in the factorization
equation (\ref{e-5-18}) is common to both planes. Taking this to
be the second factor, we have thus the additional restriction
\begin{equation}
\tilde{c}^{(1)}_z = \tilde{c}^{(2)}_z. \label{e-5-22a}
\end{equation}
This condition obviously holds when the boundary matter and gauge
groups are the same on both fixed planes, as in the HW model.

\subsection{Green-Schwarz anomaly cancellation}

Provided that the local anomaly cancellation conditions described
above are satisfied, the application of the Green-Schwarz
mechanism is straightforward. Since a Lagrangian formulation of
the theory is possible only for $n_T=0$ ($k=1$), we will
concentrate on this case. Our starting point is the factorized
anomaly polynomial
\begin{equation}
I_8 = - \frac{3}{8} \left( c_z \tr_z F^2 - \tr R^2 \right) \left(
\tilde{c}_z \tr_z F^2 - \tr R^2 \right).
\label{e-5-23}
\end{equation}
The 6D Yang-Mills and Lorentz Chern-Simons forms $\omega_{3Y,z}$
and $\omega_{3L}$ satisfy
\begin{equation}
d \omega_{3Y,z} = \tr_z F^2 ~,\qquad d \omega_{3L} = \tr R^2,
 \label{e-5-24}
\end{equation}
and their gauge and Lorentz variations are respectively given by
the  descent equations
\begin{equation}
\delta \omega_{3Y,z} = d \omega^1_{2Y,z} ~,\qquad \delta
\omega_{3L} = d \omega^1_{2L}.
 \label{e-5-25}
\end{equation}
It is not hard to see that the resulting variation of the
supergravity effective action from fermion loops can be written as
\begin{equation}
\delta \Gamma = - \xi \int \left( \tilde{c}_z \omega^1_{2Y,z} -
\omega^1_{2L} \right) \left( c_z \tr_z F^2 - \tr R^2 \right).
 \label{e-5-26}
\end{equation}
where we introduced the shorthand
\begin{equation}
\xi \equiv \frac{1}{32 (2\pi)^3}.
 \label{e-5-27}
\end{equation}

On the other hand, our Green-Schwarz term, determined by
supersymmetry considerations in Section 4 reads, completed with its gravitational part,
\begin{equation}
S_{GS} = - \frac{1}{\sqrt{2}} \int B_2 \left( v_z \tr_z F^2 - \tr
R^2 \right). \label{e-5-30}
\end{equation}
To cancel the anomalies using this term, we set all undetermined
coefficients $v_z$ to the values
\begin{equation}
v_z = c_z, \label{e-5-29}
\end{equation}
we endow $B_2$ with the anomalous gauge/Lorentz transformation law
\begin{equation}
\delta B_2 = - \sqrt{2}\xi \left( \tilde{c}_z \omega^1_{2Y,z} -
\omega^1_{2L} \right) \label{e-5-31}
\end{equation}
and we appropriately modify its field strength so that it remains
gauge/Lorentz invariant, that is, we set
\begin{equation}
G_3 = d B_2 + \sqrt{2}\xi \left( \tilde{c}_z \omega_{3Y,z} -
\omega_{3L} \right). \label{e-5-32}
\end{equation}
After these modifications, it is readily seen that the anomalous
variation of $S_{GS}$ under gauge/Lorentz transformations of $B_2$
is equal and opposite from that of $\Gamma$, yielding the desired
anomaly cancellation,
\begin{equation}
\delta \left( \Gamma + S_{GS} \right) = 0,
 \label{e-5-33}
\end{equation}
as required. It is expected that anomaly cancellation will still
be possible even in the presence of more tensor multiplets by an
appropriate modification of the Green-Schwarz
mechanism~\cite{Sagnotti:1992qw}.

\subsection{Global anomaly cancellation and non-perturbative anomalies}

Apart from the local cancellation of anomalies on the orbifold
fixed points, the anomalies of the theory must also cancel
globally on our orbifold. Labelling the two orbifold fixed planes
and all quantities pertaining to them by an index $i$ taking the
values $1$ ($x_7=0$) and $2$ ($x_7=\pi R$), we find that
cancellation of the irreducible $\tr R^4$ part of the
gravitational anomaly requires that
\begin{equation}
\sum_{i=1}^2\left( n^{(i)}_{H} + 29 n^{(i)}_{T} - n^{(i)}_{V}
\right)= 243 - N \label{e-5-34}
\end{equation}
Provided that this holds, the remaining gauge, gravitational and
mixed anomalies can cancel, in the case $n_T=0$, when the total
anomaly polynomial factorizes as
\begin{equation}
I_8^{(1)} + I_8^{(2)} = - \frac{3}{4} \left( \sum_i u^{(i)}_z
\tr_z F^{(i)2} - \tr R^2 \right) \left( \sum_i \tilde{u}^{(i)}_z
\tr_z F^{(i)2} - \tr R^2 \right). \label{e-5-35}
\end{equation}
where $u^{(i)}_z$ and $\tilde{u}^{(i)}_z$ are some numerical
constants. In the case $n_T>0$, this factorization condition need
not be satisfied; the anomaly can cancel by the generalized
Green-Schwarz mechanism.

One final issue concerns non-perturbative anomalies of the type
first discovered by Witten~\cite{Witten:fp} for the case of an odd
number of 4D Weyl fermions coupled to an $SU(2)$ gauge field. In
the present case, such anomalies may appear when one of the gauge
group factors $\mathcal{G}_z$ has non-trivial sixth homotopy group
since this implies that fermionic path integrals in certain
representations may pick up phase factors under ``large'' gauge
transformations and are therefore ill-defined. Among all possible
simple groups, the only one that can give rise to such anomalies
is $G_2$ due to the fact that $\pi_6 (G_2) = \mathbb{Z}_3$. In
such a case, the condition for the absence of such anomalies is
determined by the ``$\textrm{mod } 3$'' Atiyah-Singer index
theorem.

\section{The complete bulk-boundary action}
\label{sec-6}

We are now in a position to write down the locally supersymmetric
action resulting from the combination of the bulk action as well
as the various boundary contributions constructed above. The full
action of our theory reads
\begin{equation}
S = \int d^7x \mathcal{L}_7 + \int d^6 x \mathcal{L}_6^{(1)} +
\int d^6 x \mathcal{L}_6^{(2)}, \label{e-6-1}
\end{equation}
where $\mathcal{L}_7$ is the bulk Lagrangian given by
(\ref{e-2-34}), while $\mathcal{L}_{6}^{(i)}$ denotes the boundary
Lagrangian localized at the fixed plane $i$. For the fixed plane
at $x_7=0$, we have {\allowdisplaybreaks
\begin{eqnarray}
\label{e-6-2}
e^{-1} \mathcal{L}_{6}^{(1)} &=& - \frac{1}{4} e^{-\phi} c_z \tr_z
\left( F_{\mu\nu} F^{\mu\nu} \right) - e^\phi c_z \tr_z \left(
\bar{\lambda}^i \Gamma^\mu \mathcal{D}_\mu \lambda_i \right) -
g_{\alpha\beta}(\varphi) \mathcal{D}_\mu \varphi^\alpha
\mathcal{D}^\mu \varphi^\beta \nn \\ &&
 - \bar{\zeta}^a \Gamma^\mu
\mathcal{D}_\mu \zeta_a
- \frac{1}{12 \sqrt{2}} e^\phi \left[ - e^\phi c_z \tr_z \left(
\bar{\lambda}^i \Gamma^{\mu\nu\rho} \lambda_i \right) +
\bar{\zeta}^a \Gamma^{\nu\rho\sigma} \zeta_a \right]
G_{\mu\nu\rho}
\nn\\
&&- \frac{1}{2} c_z \tr_z \left( \bar{\psi}^i_\mu \Gamma^{\nu\rho}
\Gamma^\mu \lambda_i F_{\mu\nu} \right) - \frac{1}{2} c_z \tr_z
\left( \bar{\lambda}^i \Gamma^{\mu\nu} \chi_i F_{\mu\nu} \right) +
2 \bar{\psi}^i_\mu \Gamma^\nu \Gamma^\mu \mathcal{D}_\nu
\varphi^\alpha V_{\alpha a i} \zeta^a
\nn\\
&&+ e^\phi \tr_z \left( \C_{ij} \bar{\psi}_\mu^i \Gamma^\mu
\lambda^j \right) + e^\phi \tr_z \left( \C_{ij} \bar{\chi}^i
\lambda^j \right) - 4 e^\phi \tr_z \left( \bar{\lambda}_i \zeta_a
V_\alpha^{\phantom{\alpha} a i} \xi^\alpha \right) \nn \\ &&
- \frac{1}{2}
e^\phi c_z^{-1} \tr_z \left( \C_{ij} \C^{ij} \right)- \frac{1}{8 \sqrt{2}} e^{-1}
\epsilon^{\mu\nu\rho\sigma\tau\upsilon}  B_{\mu\nu} c_z \tr_z
\left(
 F_{\rho\sigma} F_{\tau\upsilon} \right) \nn \\ &&
-\frac{\sqrt{2}}{3} e^{\phi} C. 
\end{eqnarray}
}
On the bulk, the action (\ref{e-6-1}) is invariant under the SUSY
transformations listed in (\ref{e-2-35}). On a given boundary, the
action is invariant under the SUSY transformations obtained by
combining the transformations of the boundary theory of the
surviving bulk fields with those of the additional boundary
multiplets. These transformations are listed below.
{\allowdisplaybreaks
\begin{eqnarray}
\label{e-6-3}
\delta e^a_{\phantom{A} \mu} &=& \frac{1}{2} \bar{\epsilon}^i
\Gamma^a \psi_{\mu i},
\nn\\
\delta \phi &=& \frac{1}{2} \bar{\epsilon}^i \chi_i,
\nn\\
\delta B_{\mu\nu} &=& e^{-\phi} \left( - \frac{1}{\sqrt{2}}
\bar{\epsilon}^i \Gamma_{[\mu} \psi_{\nu] i} + \frac{1}{2
\sqrt{2}} \bar{\epsilon}^i \Gamma_{\mu\nu} \chi_i \right),
\nn\\
\delta \psi_{\mu i} &=& D_\mu \epsilon_i - \frac{1}{24 \sqrt{2}}
e^\phi \Gamma^{\nu\rho\sigma} \Gamma_\mu G_{\nu\rho\sigma}
\epsilon_i,
\nn\\
\delta \chi_i &=& \frac{1}{2} \Gamma^\mu \partial_\mu \phi
\epsilon_i - \frac{1}{12 \sqrt{2}} e^\phi \Gamma^{\mu\nu\rho}
G_{\mu\nu\rho} \epsilon_i,
\\
\delta A_\mu &=& e^\phi \bar{\epsilon}^i \Gamma_\mu \lambda_i,
\nn\\
\delta \lambda_i &=& - \frac{1}{4} e^{-\phi} \Gamma^{\mu\nu}
\epsilon_i F_{\mu\nu} - \frac{1}{2} c_z^{-1} \C_{ij} \epsilon^j,
\nn\\
\delta \varphi^\alpha &=& V^\alpha_{\phantom{\alpha} a i}
\bar{\zeta}^a \epsilon^i,
\nn\\
\delta \zeta^a &=& V_\alpha^{\phantom{\alpha} a i} \Gamma^\mu
\mathcal{D}_\mu \varphi^\alpha \epsilon^i. \nn 
\end{eqnarray}
}

The Lagrangian (\ref{e-6-2}) contains all couplings for the case of $D=6$, $\mathcal{N}=(0,1)$ supergravity coupled to one tensor multiplet, $n_V$ vector multiplets and $n_H$ hypermultiplets found in \cite{Nishino:1997ff,riccioni}. Among the various interaction terms, a particularly important role is played by the
Green-Schwarz term $B_2 \wedge \tr F^2$ (and its gravitational counterpart $B_2 \wedge \tr R^2$) that cancel the anomalies of this theory. The 7-dimensional origin of these couplings is now clarified: they appear as a consequence of dualizing the 3--form theory on a manifold with boundary.

\section{Conclusions}

In the present work, we discussed brane worlds in
seven-dimensional minimal ${\cal{N}}=2$ Yang-Mills-Einstein
supergravity. The bulk theory is the one of
\cite{Bergshoeff:1985mr}, where $N$ abelian vector multiplets are
coupled to pure 7D supergravity. In our construction we have
employed the 3--form version of \cite{Townsend:1983kk} from where
the 2--form version is obtained by Poincar\'e duality. The latter
produces surface terms, thrown  away usually, which we keep
however, as the seven-dimensional vacuum spacetime of the theory
is of the form $M^6\times {\bf S}^1/\mathbb{Z}_2$ and, thus, it
has boundaries at the fixed points of the $\mathbb{Z}_2$ action.
These surface terms are nothing else than Green-Schwarz terms,
necessary  for the cancellation of anomalies in the boundary 6D
theory. This should be anticipated with the 3--form version of the
7D theory~\cite{Gherghetta:2002xf,Gherghetta:2002nq}  and the HW
case~\cite{Horava:1995qa,Horava:1996ma}, where the existence of a
bulk Chern-Simons term is crucial for the anomaly cancellation.
This is also the case for 5D theories as well.  Here there is no
bulk Chern-Simons term to participate in the anomaly cancellation
\cite{Barbieri:2002ic}, but instead, there exists a boundary
Green-Schwarz term to do this.

The $3N$ scalars of the 7D theory parametrize the coset
$SO(N,3)/SO(N)\times SO(3)$ and an appropriate  subgroup $G\subset
SO(N,3)$ of the isometry group $SO(N,3)$ can be gauged. In that
case, supersymmetry is maintained by the introduction of a
potential, given in (\ref{pot}), for the scalars. Contrary to the
3--form version, the 2--form version of the 7D minimal
${\cal{N}}=2$ supergravity admits a 7D Minkowski vacuum. We may
then compactify the theory down to six dimensions on the orbifold
${\bf S}^1/\mathbb{Z}_2$. The resulting effective 6D theory
contains in its massless sector a gravity multiplet, a tensor
multiplet and $N+1$ vector multiplets. As this spectrum is
anomalous, extra matter fields are needed at the fixed planes to
cancel the anomalies. This matter fields can be either 6D vector,
hyper or tensor multiplets. We are not considering here extra
tensor multiplets as there is no Lagrangian in that case and we
left this possibility for future work. Here, the matter at the
boundaries necessary to cancel the anomalies comes in the form of
vector and hypers.
Following~\cite{Nishino:1997ff,riccioni,Nishino:dc,Randjbar-Daemi:1985wc,Randjbar-Daemi:2004qr},
we considered without loss of generality the case in which the
scalars parametrize the symmetric space $Sp(n_H,1)/Sp(n_H)\times
Sp(1)$, whose isometry group is  $Sp(n_H,1)$. The latter is a
global symmetry of the supergravity theory and we  gauged then its
maximal compact subgroup $Sp(n_H)\times Sp(1)$. All the results
obtained here for this particular choice of the gauge group can
easily be generalized to other quaternionic hyperscalar spaces. It
should be noted that the hypermultiplets are not neutral as
in~\cite{Gherghetta:2002xf,Gherghetta:2002nq}, but rather charged
under the gauge group. In addition, both the vector and
hypermultiplets are coupled to the bulk fields and their couplings
have been specified by supersymmetry.

We should also mention that, since the brane worlds constructed
here are six dimensional, compactification down to four dimensions
is needed. Such a compactification which involves an
$\mathbf{S}^2$, much like the monopole compactification~\cite{SS},
will be presented elsewhere.

\[ \]
{\bf Acknowlegements} We wish to thank T. Gherghetta for interesting
discussions. S.D.A. also wishes to thank the CERN-TH Division
for hospitality during the late stages of this work. This work is
partially supported by the EPAN projects, 68/664 Heraclitus and
Pythagoras, and the Protagoras-8198 NTUA project.

\appendix

\section{Conventions}
\label{appa}

We closely follow the sign and normalization conventions used
in~\cite{Gherghetta:2002xf,Gherghetta:2002nq}. Our spacetime
metric has the ``mostly-plus'' signature $(-,+,\ldots,+)$. The 6D
$8 \times 8$ matrices $\Gamma^a$ satisfy the Clifford algebra
\begin{equation} \{
\Gamma^a , \Gamma^b \} = 2 \eta^{ab} ~,\qquad a,b=0,\ldots,5,
\label{e-a-1}
\end{equation}
while the 7D gamma matrices are $\Gamma^A = (\Gamma^a,\Gamma_7)$
with the latter given by
\begin{equation} \Gamma_7 = \Gamma^0 \ldots \Gamma^5,
\label{e-a-2}
\end{equation}
so that $(\Gamma_7)^2=1$. The 6D gamma matrices satisfy the
duality relation
\begin{equation}
\Gamma^{\mu_1 \ldots \mu_n} = \frac{ (-1)^{[n/2]} }{(6-n)!}
\epsilon^{\mu_1 \ldots \mu_n \mu_{n+1} \ldots \mu_6}
\Gamma_{\mu_{n+1} \ldots \mu_6} \Gamma_7.
\label{a-3}
\end{equation}
Our spinors satisfy the symplectic Majorana condition
\begin{equation}
\chi^i = \epsilon^{ij} \bar{\chi}^T_j ~,\qquad \bar{\chi}_i =
\chi^{i\dag} \Gamma_0 \label{e-a-4}
\end{equation}
where $i,j=1,2$ are $Sp(1)$ R-symmetry indices and $\epsilon_{ij}$
is the $Sp(1)$--invariant tensor, defined as
\begin{equation}
\epsilon_{ij} = \epsilon^{ij} = \left( \begin{array}{cc} 0 & 1
\\ -1 & 0 \end{array} \right).
\label{e-a-5}
\end{equation}
This tensor is used for raising and lowering $Sp(1)$ indices
according to the standard NW-SE convention,
\begin{equation} \chi^i
= \epsilon^{ij} \chi_j ~,\qquad \chi_i = \chi^j \epsilon_{ji}
. \label{e-a-6}
\end{equation}
The same convention is employed in the contraction of $Sp(1)$
indices in all spinor inner products,
\begin{equation}
\bar{\chi} \Gamma^{M_1 \ldots M_n} \psi \equiv \bar{\chi}^i
\Gamma^{M_1 \ldots M_n} \psi_i. \label{e-a-7}
\end{equation}
We also note that the following identities hold
\begin{equation}
\bar{\chi}^i \Gamma^{M_1 \ldots M_n} \psi^j = (-1)^{n+1}
\bar{\psi}^j \Gamma^{M_n \ldots M_1} \chi^i. \label{e-a-8}
\end{equation}
and
\begin{equation}
\bar{\chi} \Gamma^{M_1 \ldots M_n} \psi =
(-1)^n \bar{\psi} \Gamma^{M_1 \ldots M_n} \chi. \label{e-a-9}
\end{equation}
In 6 dimensions, a symplectic Majorana spinor decomposes into
positive- and negative-chirality parts according to $\chi^i =
\chi^{i+} + \chi^{i-}$, where $\chi^{i\pm}$ are symplectic
Majorana-Weyl spinors satisfying $\Gamma_7 \chi^{i\pm} = \pm
\chi^{i\pm}$.

\section{Anomaly polynomials}
\label{appb}

In the discussion of anomaly cancellation, we will use the
following normalization for the 8--form anomaly polynomials,
\begin{eqnarray}
\label{e-b-1}
I^{1/2}_{8} (F) &=& - \tr F^4, \nn\\
I^{1/2}_{8} (R) &=& - \frac{1}{240} \tr R^4 - \frac{1}{192} (\tr R^2)^2, \nn\\
I^{1/2}_{8} (F,R) &=& \frac{1}{4} \tr R^2 \tr F^2, \\
I^{3/2}_{8} (R) &=& - \frac{49}{48} \tr R^4 + \frac{43}{192} (\tr R^2)^2 ,\nn\\
I^A_{8} (R) &=& - \frac{7}{60} \tr R^4 + \frac{1}{24} (\tr R^2)^2. \nn
\end{eqnarray}
where the superscripts $1/2$, $3/2$ and $A$ refer to a spin $1/2$
fermion, a spin $3/2$ fermion and a 2--form antisymmetric tensor
potential respectively. The above anomaly polynomials correspond
to Weyl spinors of positive chirality and 2--form potentials with
self-dual field strengths. For a Majorana-Weyl spinor, one needs
to include a factor of $\frac{1}{2}$, while for a
negative-chirality spinor (or an anti-self-dual field strength)
the sign of the anomaly must be reversed.


\begin{thebibliography}{x}

\bibitem{Arkani-Hamed:2001is}
N.~Arkani-Hamed, A.~G.~Cohen and H.~Georgi,
Phys.\ Lett.\ B {\bf 516}, 395 (2001) [arXiv:hep-th/0103135].

\bibitem{Arkani-Hamed:1998rs}
N.~Arkani-Hamed, S.~Dimopoulos and G.~R.~Dvali,
Phys.\ Lett.\ B {\bf 429}, 263 (1998) [arXiv:hep-ph/9803315].

\bibitem{Randall:1999ee}
L.~Randall and R.~Sundrum,
Phys.\ Rev.\ Lett.\  {\bf 83}, 3370 (1999) [arXiv:hep-ph/9905221].

\bibitem{Randall:1999vf}
L.~Randall and R.~Sundrum,
Phys.\ Rev.\ Lett.\  {\bf 83}, 4690 (1999) [arXiv:hep-th/9906064].

\bibitem{Scrucca:2004jn}
C.~A.~Scrucca and M.~Serone,
arXiv:hep-th/0403163.

\bibitem{Horava:1995qa}
P.~Ho\v rava and E.~Witten,
Nucl.\ Phys.\ B {\bf 460}, 506 (1996) [arXiv:hep-th/9510209].

\bibitem{Horava:1996ma}
P.~Ho\v rava and E.~Witten,
Nucl.\ Phys.\ B {\bf 475}, 94 (1996) [arXiv:hep-th/9603142].

\bibitem{Kehagias:1999ju}
A.~Kehagias,
Phys.\ Lett.\ B {\bf 469}, 123 (1999)
[arXiv:hep-th/9906204].

\bibitem{Kehagias:2000au}
A.~Kehagias and K.~Tamvakis,
Phys.\ Lett.\ B {\bf 504}, 38 (2001)
[arXiv:hep-th/0010112].

\bibitem{GrootNibbelink:2002qp}
S.~G. Nibbelink, H.~P.~Nilles and M.~Olechowski,
Nucl.\ Phys.\ B {\bf 640}, 171 (2002)
[arXiv:hep-th/0205012].

\bibitem{Gmeiner:2002ab}
F.~Gmeiner, J.~E.~Kim, H.~M.~Lee and H.~P.~Nilles,
arXiv:hep-th/0205149.

\bibitem{Gmeiner:2002es}
F.~Gmeiner, S.~G. Nibbelink, H.~P.~Nilles, M.~Olechowski and
M.~G.~A.~Walter,
Nucl.\ Phys.\ B {\bf 648}, 35 (2003)
[arXiv:hep-th/0208146].

\bibitem{Alvarez-Gaume:1983ig}
L.~Alvarez-Gaum\'e and E.~Witten,
Nucl.\ Phys.\ B {\bf 234}, 269 (1984).

\bibitem{Salam:1985mi}
A.~Salam and E.~Sezgin,
Phys.\ Scripta {\bf 32}, 283 (1985).

\bibitem{Bergshoeff:1986hv}
E.~Bergshoeff, T.~W.~Kephart, A.~Salam and E.~Sezgin,
Mod.\ Phys.\ Lett.\ A {\bf 1}, 267 (1986).

\bibitem{Dobrescu:2001ae}
B.~A.~Dobrescu and E.~Poppitz,
Phys.\ Rev.\ Lett.\  {\bf 87}, 031801 (2001)
[arXiv:hep-ph/0102010].

\bibitem{Appelquist:2002ft}
T.~Appelquist, B.~A.~Dobrescu, E.~Ponton and H.~U.~Yee,
Phys.\ Rev.\ D {\bf 65}, 105019 (2002) [arXiv:hep-ph/0201131].

\bibitem{Gherghetta:2002xf}
T.~Gherghetta and A.~Kehagias,
Phys.\ Rev.\ Lett.\  {\bf 90}, 101601 (2003)
[arXiv:hep-th/0211019].

\bibitem{Gherghetta:2002nq}
T.~Gherghetta and A.~Kehagias,
Phys.\ Rev.\ D {\bf 68}, 065019 (2003) [arXiv:hep-th/0212060].

\bibitem{GrootNibbelink:2003gd}
S.~G. Nibbelink,
JHEP {\bf 0307}, 011 (2003)
[arXiv:hep-th/0305139].

\bibitem{Nibbelink:2003rc}
S.~G.~Nibbelink, M.~Hillenbach, T.~Kobayashi and M.~G.~A.~Walter,
Phys.\ Rev.\ D {\bf 69}, 046001 (2004)
[arXiv:hep-th/0308076].

\bibitem{Gorbatov:2001pw}
E.~Gorbatov, V.~S.~Kaplunovsky, J.~Sonnenschein, S.~Theisen and
S.~Yankielowicz,
JHEP {\bf 0205}, 015 (2002) [arXiv:hep-th/0108135].

\bibitem{Kaplunovsky:1999ia}
V.~Kaplunovsky, J.~Sonnenschein, S.~Theisen and S.~Yankielowicz,
Nucl.\ Phys.\ B {\bf 590}, 123 (2000) [arXiv:hep-th/9912144].

\bibitem{Faux:2000sp}
M.~Faux, D.~Lust and B.~A.~Ovrut,
Int.\ J.\ Mod.\ Phys.\ A {\bf 18}, 3273 (2003)
[arXiv:hep-th/0010087].

\bibitem{Maldacena:1997re}
J.~M.~Maldacena,
Adv.\ Theor.\ Math.\ Phys.\  {\bf 2}, 231 (1998) [Int.\ J.\
Theor.\ Phys.\  {\bf 38}, 1113 (1999)] [arXiv:hep-th/9711200].

\bibitem{Ferrara:1998vf}
S.~Ferrara, A.~Kehagias, H.~Partouche and A.~Zaffaroni,
Phys.\ Lett.\ B {\bf 431}, 42 (1998) [arXiv:hep-th/9803109].

\bibitem{Kehagias:1998gn}
A.~Kehagias,
Phys.\ Lett.\ B {\bf 435}, 337 (1998)
[arXiv:hep-th/9805131].

\bibitem{Witten:1995ex}
E.~Witten,
Nucl.\ Phys.\ B {\bf 443}, 85 (1995) [arXiv:hep-th/9503124].

\bibitem{Mezincescu:ta}
L.~Mezincescu, P.~K.~Townsend and P.~van Nieuwenhuizen,
Phys.\ Lett.\ B {\bf 143}, 384 (1984).

\bibitem{Townsend:1983kk}
P.~K.~Townsend and P.~van Nieuwenhuizen,
Phys.\ Lett.\ B {\bf 125}, 41 (1983).

\bibitem{Salam:1983fa}
A.~Salam and E.~Sezgin,
Phys.\ Lett.\ B {\bf 126}, 295 (1983).

\bibitem{Han:1985ku}
S.~K.~Han, I.~G.~Koh and H.~W.~Lee,
Phys.\ Rev.\ D {\bf 32}, 3190 (1985).

\bibitem{Giani:dw}
F.~Giani, M.~Pernici and P.~van Nieuwenhuizen,
Phys.\ Rev.\ D {\bf 30}, 1680 (1984).

\bibitem{Bergshoeff:1985mr}
E.~Bergshoeff, I.~G.~Koh and E.~Sezgin,
Phys.\ Rev.\ D {\bf 32}, 1353 (1985).

\bibitem{Park:id}
Y.~J.~Park,
Phys.\ Rev.\ D {\bf 38}, 1087 (1988).

\bibitem{Bagger:2002rw}
J.~Bagger and D.~V.~Belyaev,
Phys.\ Rev.\ D {\bf 67}, 025004 (2003)
[arXiv:hep-th/0206024].

\bibitem{Cremmer:1979up}
E.~Cremmer and B.~Julia,
Nucl.\ Phys.\ B {\bf 159}, 141 (1979).

\bibitem{Nicolai:1980td}
H.~Nicolai and P.~K.~Townsend,
Phys.\ Lett.\ B {\bf 98}, 257 (1981).

\bibitem{Conrad:1997ww}
J.~O.~Conrad,
Phys.\ Lett.\ B {\bf 421}, 119 (1998)
[arXiv:hep-th/9708031].

\bibitem{Harmark:1998bs}
T.~Harmark,
Phys.\ Lett.\ B {\bf 431}, 295 (1998) [arXiv:hep-th/9802190].

\bibitem{Sagnotti:1992qw}
A.~Sagnotti,
Phys.\ Lett.\ B {\bf 294}, 196 (1992)
[arXiv:hep-th/9210127].

\bibitem{Dudas:2004ni}
E.~Dudas, T.~Gherghetta and S.~G.~Nibbelink,
arXiv:hep-th/0404094.

\bibitem{FR}
S.~Ferrara, F.~Riccioni and A.~Sagnotti,
Nucl.\ Phys.\ B {\bf 519}, 115 (1998) [arXiv:hep-th/9711059].

\bibitem{Nishino:1997ff}
H.~Nishino and E.~Sezgin,
Nucl.\ Phys.\ B {\bf 505}, 497 (1997) [arXiv:hep-th/9703075].

\bibitem{riccioni}
F.~Riccioni,
Nucl.\ Phys.\ B {\bf 605}, 245 (2001) [arXiv:hep-th/0101074].

\bibitem{Nishino:dc}
H.~Nishino and E.~Sezgin,
Nucl.\ Phys.\ B {\bf 278}, 353 (1986).

\bibitem{Randjbar-Daemi:1985wc}
S.~Randjbar-Daemi, A.~Salam, E.~Sezgin and J.~Strathdee,
Phys.\ Lett.\ B {\bf 151} (1985) 351.

\bibitem{Randjbar-Daemi:2004qr}
S.~Randjbar-Daemi and E.~Sezgin,
arXiv:hep-th/0402217.

\bibitem{Erler:1993zy}
J.~Erler,
J.\ Math.\ Phys.\  {\bf 35}, 1819 (1994) [arXiv:hep-th/9304104].

\bibitem{Schwarz:1995zw}
J.~H.~Schwarz,
Phys.\ Lett.\ B {\bf 371}, 223 (1996) [arXiv:hep-th/9512053].

\bibitem{Green:1984bx}
M.~B.~Green, J.~H.~Schwarz and P.~C.~West,
Nucl.\ Phys.\ B {\bf 254}, 327 (1985).

\bibitem{Witten:fp}
E.~Witten,
Phys.\ Lett.\ B {\bf 117}, 324 (1982).

\bibitem{Barbieri:2002ic}
R.~Barbieri, R.~Contino, P.~Creminelli, R.~Rattazzi and C.~A.~Scrucca,
Phys.\ Rev.\ D {\bf 66}, 024025 (2002)
[arXiv:hep-th/0203039].

\bibitem{SS}
A.~Salam and E.~Sezgin,
Phys.\ Lett.\ B {\bf 147}, 47 (1984).

\end{thebibliography}
\end{document}